\documentclass[notitlepage,reprint,onecolumn, amsmath,amssymb, aps,prd, nofootinbib]{revtex4-1} 

\usepackage{graphicx}
\usepackage{epsfig}
\usepackage{bm}
\usepackage{color}
\usepackage[utf8]{inputenc} 
\usepackage[T1]{fontenc}
\usepackage{xfrac}
\usepackage{booktabs} 
\usepackage{array} 
\usepackage{paralist} 
\usepackage{verbatim} 
\usepackage{slashed}

\numberwithin{equation}{section}

\definecolor{DGreen}{rgb}{0.0, 0.42, 0.24}

\newcommand{\bit}{\begin{itemize}}
\newcommand{\eit}{\end{itemize}}
\newcommand{\ben}{\begin{enumerate}}
\newcommand{\een}{\end{enumerate}}
\newcommand{\beq}{\begin{equation}}
\newcommand{\eeq}{\end{equation}}
\newcommand{\bea}{\begin{eqnarray}}
\newcommand{\eea}{\end{eqnarray}}
\newcommand{ \lsim}{\mathrel{\vcenter
   {\hbox{$<$}\nointerlineskip\hbox{$\sim$}}}}
\newcommand{ \gsim}{\mathrel{\vcenter
   {\hbox{$>$}\nointerlineskip\hbox{$\sim$}}}}
\newcommand{\gappeq}{\mathrel{\rlap {\raise.5ex\hbox{$>$}}
{\lower.5ex\hbox{$\sim$}}}}
\newcommand{\lappeq}{\mathrel{\rlap{\raise.5ex\hbox{$<$}}
{\lower.5ex\hbox{$\sim$}}}}

\newcommand{\Zslash}{ \, Z \! \! \! \! / ~ }

\newcommand{\muc}{\mu A \! \to \! eA }
\newcommand{\mucAu}{\mu {\rm Au} \to e {\rm Au} }
\newcommand{\mucAl}{\mu {\rm Al} \to e {\rm Al}}
\newcommand{\mucAuL}{\mu {\rm Au} \to e_L {\rm Au} }
\newcommand{\mucAlL}{\mu {\rm Al} \to e_L {\rm Al}}
\newcommand{\mucL}{\mu A\to e_L A}
\newcommand{\mec}{\mu \! \to \! e~ {\rm conversion}}

\newcommand{\meg}{\mu \to e \gamma}
\newcommand{\megL}{\mu \to e_L \gamma}

\newcommand{\Zme}{Z \to e^\pm \mu^\mp }

\newcommand{\meee}{\mu \to e \bar{e} e}
\newcommand{\meeeL}{\mu \to e_L \bar{e} e}

\newcommand{\LNP}{\Lambda_{LFV}}

\def\a{\alpha}
\def\b{\beta}
\def\g{\gamma}
\def\d{\delta}

\def\m{\mu}

\begin{document}

\title{Reach and complementarity of $\mu\to e$ searches}

\author{Sacha Davidson}
\email{E-mail address: s.davidson@lupm.in2p3.fr}
\affiliation{LUPM, CNRS,
Universit\'e Montpellier
Place Eugene Bataillon, F-34095 Montpellier, Cedex 5, France}
\author{Bertrand Echenard}
\email{E-mail address: echenard@caltech.edu}
\affiliation{California Institute of Technology, Pasadena, California 91125 USA}

\begin{abstract}
\noindent
In Effective Field Theory, we describe $\mu\leftrightarrow e$ flavour changing transitions using an operator basis motivated by 
experimental observables. In a six-dimensional subspace probed by $\meg$, $\meee$ and $\mu\to e$ conversion on nuclei, we derive 
constraints on the New Physics  scale from past and future experiments, illustrating the complementarity of the 
processes in an intuitive way.
\end{abstract}

\maketitle


\section{Introduction}
\label{intro}

Lepton flavor-violating contact interactions --- referred to as LFV or CLFV --- are excellent probes of physics beyond the 
standard model (see e.g.~\cite{KO,Calibbi} for reviews). The non-zero neutrino masses and mixing angles imply their existence, 
and an observation could shed light on the neutrino mass mechanism~\cite{mnu}, and even on the matter excess of our universe 
if generated via leptogenesis~\cite{Y+,PRep}. The rates could be just below the current experimental bounds, in many extensions 
of the Standard Model that introduce additional CLFV sources~\cite{Calibbi}.

CLFV searches have been conducted in a wide range of reactions, and a subset of current and anticipated experimental constraints 
are given in Table~\ref{tab:bds}. While the multitude of $\tau$ channels could contribute to identifying the nature of New 
Physics (NP), the greater sensitivity in muon sector seems more promising for discovering it. Although the number of muon processes 
is limited, the corresponding bounds are already quite restrictive, and exceptional improvements are expected in the coming years.

\begin{table}[ht]
\begin{center}
 \begin{tabular}{|l|l|l|}
 \hline
 process & current sensitivity & future \\
\hline
$\meg $ & $ < 4.2 \times 10^{-13}$(MEG~\cite{TheMEG:2016wtm})
 &$ \sim 10^{-14}$ (MEG II~\cite{MEGII}) \\
$\meee $& $ < 1.0 \times 10^{-12}$(SINDRUM~\cite{Bellgardt:1987du}) 
 & $ \sim 10^{-16}$ (Mu3e~\cite{Mu3e}) \\
$\mu$A$ \to e$A & 
$< 7 \times 10^{-13}$(SINDRUM II~\cite{Bertl:2006up}) &
 $ \sim 10^{-16}$ (COMET~\cite{COMET}, Mu2e~\cite{mu2e})  \\
  & &
  $ \sim 10^{-(18\to 20)}$ (PRISM/PRIME~\cite{PP})  \\
\hline
$\tau\to l\gamma$ & $<3.3\times 10^{-8}$(Babar)~\cite{tau1} & $ \sim 10^{-9} $(BelleII)~\cite{belle2t3l} \\
$\tau\to 3l$ & $<$ few$\times 10^{-8}$ (Belle)~\cite{tau2}
& $ \sim 10^{-9} $(BelleII)~\cite{belle2t3l} \\
$\tau\to l\pi^0$ & $<8.0\times 10^{-8}$ (Belle)~\cite{Belle:2007cio} & $ \sim 10^{-9} $ (BelleII)~\cite{belle2t3l} \\
 $\tau\to l \rho$ & $<1.2\times 10^{-8}$ (Belle)~\cite{Belle:2007cio} & $ \sim 10^{-9}$ (BelleII)~\cite{belle2t3l} \\
\hline
 \end{tabular}
 \caption{Current bounds on the branching ratios for various lepton flavour changing processes, and the expected reach 
 of upcoming experiments. \label{tab:bds} }
 \end{center}
 \end{table}

The reach and complementarity of $\meg$, $\meee$ and $\muc$ transitions have been explored from various perspectives. Numerous 
authors have investigated model preferences and predictions for correlations among CLFV observables~\cite{everyone}. An early 
model-independent analysis was performed by de Gouvea and Vogel~\cite{deGouvea}, using a simple effective Lagrangian to describe 
NP effects. The presentation of our results follows their intuitive plots. However, their approach was limited to comparing 
the reach of pairs of processes (e.g. $\meg$ {\it vs.} $\muc$) at tree level. More systematic Effective Field Theory (EFT)
studies, including more operators and some loop effects, were performed in~\cite{PSI,C+C};~\cite{PSI} focused on the sensitivity 
of the processes to a more complete operator basis, and~\cite{C+C} explored whether the proposed experimental muon program 
is necessary and sufficient to find $\mu\to e$ flavour change.

The aim of this work is to graphically illustrate the complementarity and reach of the $\meg$, $\meee$ and $\muc$ processes. We 
describe the  physics of CLFV  in an EFT perspective~\cite{Georgi,burashouches,LesHouches}, where the number of operator 
coefficients is reduced  by choosing an operator basis motivated by our observables~\cite{C+C}. We quantify complementarity 
as the degree to which observables probe different operator coefficients, and study the complementarity of observables at the 
New Physics scale because the aim is to make observations that give distinct information at $\LNP$. The coefficients are translated 
to $\LNP$ using  Renormalisation Group Equations. This study extends the analysis described in~\cite{C+C} in several ways: {we 
provide more informative plots of the current and projected experimental reaches}, and a more rigorous construction of the basis for 
the subspace of experimentally accessible operator coefficients. In addition we draw attention to the information loss in matching 
nucleons to quarks using current theoretical results. Using this formalism to study whether $\mu\to e$ processes can distinguish 
among models is an interesting question that we leave for a subsequent publication.

This paper is organized as follows: section~\ref{sec:theorie} outlines the procedure to take the data parametrized in EFT from the 
experimental scale to beyond the weak scale. Section~\ref{sec:plots} presents constraints from various experimental measurements 
and projections for future initiatives. The construction of the basis used in this work is discussed in Appendices~\ref{app:ops} 
and~\ref{app:ping}. An independent issue regarding information loss in relating $\muc$ rates to models is finally discussed in 
Appendix~\ref{app:yeeks}.


\section{Theory overview}
\label{sec:theorie}

This section gives the Lagrangian and Branching Ratios at the experimental scale, and sketches the transformation of operator 
coefficients from the experimental scale to $\LNP$ (which is described in more detail in ~\cite{C+C}).

A challenge of the EFT approach lies in the large number of operators. In the case of $\mu \to e$ flavour changing processes, about 90 
operators~\cite{C+C} are required to parametrize interactions that have $\leq 4$ Standard Model legs at low energy and are otherwise 
flavour-diagonal. The difficulty to constrain and visualize this high-dimensional space is compounded by the fact that there are (only) 
three processes with excellent sensitivity in the $\mu \to e$ sector (see Table~\ref{tab:bds}), imposing only about a dozen constraints 
on operator coefficients~\cite{DKY}. Improved theoretical calculations and additional $\muc$ measurements with different nuclear targets 
could increase this number to $\sim 20$ independent constraints~\cite{DKY}. Determining all EFT coefficients appears therefore a 
daunting task. 

This manuscript takes a different perspective, following~\cite{C+C}. Since there are three processes with excellent experimental sensitivity,
we restrict to the (12-dimensional) subspace of operator coefficients probed by $\meg,\meee$, and Spin Independent\footnote{We leave Spin Dependent 
conversion~\cite{CDK,DKS,DKY} and other targets~\cite{DKY} for future work.} $\mu Al \to e Al$ and $\mu Au \to e Au$.
The dimension of the subspace can be further reduced by half since the operator coefficients can be labelled by the helicity (or chirality) 
of the outgoing relativistic electron, and the results are very similar for either $e_L$ or $e_R$, which do not interfere. Restricting 
the analysis to the subspace corresponding to an outgoing $e_L$ in the bilinear with a muon, the three muon processes can be described at the experimental scale 
($\sim m_\mu$) by the following effective Lagrangian~\cite{KO}:
\bea
\d {\cal L} &=& \frac{1}{\LNP^2}{\Big [}C_{D} (m_\mu \overline{e} \sigma^{\a\b}P_{R} \mu) F_{\a\b}+ C_{S} (\overline{e} P_R \mu ) 
(\overline{e} P_R e ) + C_{VR} (\overline{e} \gamma^\a P_L \mu ) (\overline{e} \gamma_\a P_R e )
\nonumber\\
&&\phantom{\frac{1}{\LNP^2}{\Big [}} + C_{VL} (\overline{e} \gamma^\a P_L \mu ) (\overline{e} \gamma_\a P_L e ) + C_{Alight} {\cal O}_{Alight} + C_{Aheavy\perp} {\cal O}_{Aheavy\perp} {\Big ]} 
\label{Lag1}
\eea
where $\LNP$ is the New Physics  scale, and the dimensionless coefficients $\{C_Z\}$ are lined up in a vector $\vec{C}$ normalised to 1 at the experimental scale. 
The first term of this Lagrangian is a dipole operator mediating $\meg$ and contributing to both $\meee$ and $\muc$. The 
next three contact operators contribute to $\meee$, while ${\cal O}_{Alight} $ is a combination of operators probed by light 
muon conversion targets such as Ti or Al, and $ {\cal O}_{Aheavy\perp}$ is an orthogonal combination probed by heavy targets 
such as Au.
An approximate expression for  ${\cal O}_{Alight}$ at the experimental scale  is given in Eq. (\ref{A2}), and for  ${\cal O}_{Alight}$ and for  ${\cal O}_{Aheavy\perp}$ at 2 GeV in Eq. (\ref{A14}).
We take Au and Al as prototypical ``heavy'' and ``light'' targets since Au was used by the 
SINDRUM experiment~\cite{Bertl:2006up}, and Al will be used by the upcoming COMET~\cite{COMET} and Mu2e~\cite{mu2e} 
experiments, in addition to resembling Ti used in the past~\cite{Bertl:2006up}. 

The constraint on the dipole operator from $\meg$ is given by: 
 \bea
BR(\megL)& =& 384 \pi^2 \frac{v^4}{\LNP^4} |\vec{C}\cdot \hat{e}_{D R}|^2 <B_{\meg}^{expt} = 4.2\times 10^{-13}
\label{BRmeg}
\eea
where we introduced unit vectors $\hat{e}_A$ which select coefficients $C_A$ in the six-dimensional subspace. The four-lepton 
operators have negligeable interference  in $\meee$ since the electrons are relativistic ($\approx$ chiral), setting the 
three following constraints:
\bea
  BR(\mu \to e_L\overline{e_R} e_L) &=& \frac{v^4}{\LNP^4} \frac{|\vec{C}\cdot \hat{e}_{S}|^2}{8}  \leq B^{expt}_{\meee}= 10^{-12} \label{BRmeee}\\
  BR(\mu \to e_L\overline{e_L} e_L) &=& \frac{v^4}{\LNP^4}\left[2| \vec{C}\cdot \hat{e}_{VL} + 4e\vec{C}\cdot \hat{e}_{D}|^2
     + e^2(32 \ln\frac{m_\mu}{m_e} -68) |\vec{C}\cdot \hat{e}_{D}|^2\right] \leq B_{\meee}^{expt} \nonumber\\
  BR(\mu \to e_L\overline{e_R} e_R) &=& \frac{v^4}{\LNP^4}\left[| \vec{C}\cdot \hat{e}_{VR} + 4e\vec{C}\cdot \hat{e}_{D}|^2
     + e^2(32 \ln\frac{m_\mu}{m_e} -68) |\vec{C}\cdot \hat{e}_{D}|^2\right] \leq B_{\meee}^{expt}~~~.
\nonumber
\eea
These can conveniently be summarised as
\beq
  BR(\meeeL) = \frac{v^4}{\LNP^4} \vec{C} ^\dagger \bm{R}_{\meeeL} \vec{C}  \leq B_{\meee}^{expt}
\eeq
where the matrix {\bf R}$_{\meeeL}$ is proportional to the inverse covariance matrix for $\meee$, given in eqn (\ref{Rmeee}). Similarly, 
the Conversion Ratios for $\muc$ can be written 
\bea 
  CR(\mu Al \to e_L Al) &=&\frac{v^4}{\LNP^4} \vec{C} ^\dagger \bm{R}_{\mucAlL} \vec{C}\leq B_{\mu Al \to e Al}^{expt}\nonumber\\
  CR(\mu Au \to e_L Au) &=&\frac{v^4}{\LNP^4} \vec{C} ^\dagger \bm{R}_{\mucAuL} \vec{C}\leq B_{\mu Au \to e Au}^{expt}\label{BRmucL}
 \eea
where the {\bf R} matrices are given in eqn (\ref{preR2G}). These expressions justify a posteriori the basis in 
eqn (\ref{Lag1}), chosen to be orthogonal, intuitive, and correspond closely to the coefficient combinations probed 
by observables. The eigenvectors of the covariance matrix could be another basis choice, discussed briefly in Appendix~\ref{app:V}.

From the Lagrangian (\ref{Lag1}), one can easily deduce that the three processes are complementary at the experimental scale: 
four-fermion interactions with leptons only contribute to $\meee$, interactions with strongly-interacting particles only 
contribute to $\muc$ (the complementarity between heavy or light targets is discussed at the end of Appendix~\ref{app:ops}), 
while the dipole contributes to all processes. The complementarity of two processes can be interpreted geometrically as the 
misalignment between the corresponding vectors in coefficient space, defined as the angle $\eta$ between the two vectors. In terms 
of {\bf R} matrices, the complementarity between processes $A$ and $B$ can be expressed as:
\beq
  \cos \eta \sim \frac{{\rm Tr} {\Big [} \bm{R}_{A} \bm{R}_{B} {\Big ]}} {{\rm Tr} {\Big [} \bm{R}_{A} {\Big ]} {\rm Tr} {\Big [}\bm{R}_{B} {\Big ]}} ~~~,
  \label{TrR}
\eeq
which vanishes for perfectly complementary observables, and is equal to one when they contain the same information. The basis in 
eqn (\ref{Lag1}) was chosen to be perfectly complementary, i.e. orthogonal, at the experimental scale. 

In the following, we adopt a different approach to illustrate the complementarity between processes. Instead of using the geometric 
measure defined above, we show that each process gives independent information about the operator coefficients by plotting the 
corresponding reach separately. The measured rates can then be combined to identify a point in parameter space.

The degree of complementarity can be evaluated at $\Lambda_{LFV}$ by translating the coefficients in eqn (\ref{Lag1}) from the 
experimental scale to $\LNP$. Modifying the scale amounts to changing the separation between lower energy loop effects that are explicitly 
calculated, and higher energy loops that are implicitly resummed into the coupling constants. At the experimental scale, all the loops via 
which a New Physics model contributes to an observable are in the operator coefficients, and the rate is straightforward to calculate. On the 
other hand, the operator coefficients at the heavy LFV scale are easily derived from a New Physics model, but loops must be calculated to 
evaluate experimental quantities. So the low-energy operator coefficients of eqn (~\ref{Lag1}) can be transformed to the LFV scale 
$\LNP$ via the Renormalisation Group Equations (RGEs)~\cite{PSI,JMT}, which peel off the SM loops in a leading log expansion. 
Solving the RGEs perturbatively, and modifying the EFT with scale to account for the changing particle content, allows us to write
\beq
  \vec{C}(m_\mu)= \bm{G}^T(\LNP,m_\mu)\vec{C}(\LNP) ~~~.
\label{pt5}
\eeq
The matrix {\bf G} is similar to that given in~\cite{C+C}. We neglect loop effects in the EFT of nucleons and pions and match at 2 GeV onto 
a QCD-invariant EFT with gluons and five flavours of quarks (see Appendix~\ref{app:ops} and Table~\ref{tab:G}). The leading log QED and 
QCD effects are included up to $m_W$~\cite{PSI,Roma}, where the coefficients are matched (at tree level in the lower-energy EFT) onto 
dimension six SMEFT operators, augmented by the dimension eight scalar operator~\cite{Murphy, autredim8} corresponding to $\hat{e}_{S}$, 
which could be relevant~\cite{powercount}. We neglected CKM angles in matching and some other relevant SMEFT operators of dimension eight, 
and stress that the running from $m_W\to \LNP$ is {not} included. 

The Branching Ratios in terms of coefficients at $\LNP$ can be expressed as:
\bea
  BR(\mu\to e_LX) &= & \vec{C} ^\dagger(m_W) \bm{G}^* (m_W,m_\mu) \bm{R}_{\mu\to e_L X} (m_\mu) \bm{G}^T (m_W,m_\mu) \vec{C}(m_W)
\label{RmegLam}
\eea
where the matrix {\bf G} is {\it not} unitary, and does not preserve the orthonormality of the basis since SM loops and matching can
change the normalisation and direction of the vectors $\{\hat{e}_A\}$. This is expected since distinct observations at low energy 
can measure the same high-scale NP coefficients. 
The changing modulus of the basis vectors is simple to calculate and include, and affects the reach. The changes in direction can 
affect the complementarity of processes if the vectors become more or less aligned (see Appendix ~\ref{app:ping}).


\section{Illustrating experimental constraints}
\label{sec:plots}

In this section, we illustrate the constraints on New Physics from current and future $\mu \to e$ searches, and show how these results 
can be combined to identify the allowed region of coefficient space. We parametrize the coefficient space with spherical 
coordinates~\cite{sphcoord} (Table~\ref{tab:angles}) assuming that the vector of coefficients $\vec{C}$ is normalised to unity at the 
experimental scale. The reach of the various experiments in $\LNP$ can be calculated as a function of these angles and the branching 
ratios given in eqn~\ref{BRBE}. We stress that we are showing (projected) exclusion curves, as opposed to ``one-at-a-time'' bounds, 
since our EFT formulation should account for potential cancellations in the theoretical rate.

In deriving this parametrization, we approximate the operator coefficients as real numbers. This familiar simplification reduces our 
coefficient space from six complex to six real dimensions, replacing relative phases between interfering coefficients with a relative sign. 
Furthermore, we focus on a four-dimensional subspace, corresponding approximately to the four processes we examine, by suppressing two 
of the three four-lepton directions (the four-lepton operators can be distinguished by measuring the angular distribution in 
$\meee$~\cite{Okadameee}). The direction $\vec{e}_S$ associated to the scalar four lepton operator interferes with none of the other operators 
and receives negligible loop corrections, so it is complementary by inspection. We also neglect a linear combination of 
the vector four-lepton directions $\vec{e}_{VR}$ and $\vec{e}_{VL}$, since their contributions to $\meee$ have similar form. A judicious 
choice ensures the approximate orthogonality of the remaining four basis vectors. The full details are given in Appendix~\ref{app:ping}. Modulo these approximations, the  parametrisation describes the experimentally constrainable space, so we  now plot various slices through the excluded region  to illustrate its shape.  

\begin{table}[htb]
\begin{center}
 \begin{tabular}{|l|l|}\hline
  $\vec{C}\cdot \vec{e}_D$             & $|\vec{e}_D|\cos \theta_D $\\
  $\vec{C}\cdot \vec{e}_{S }$          & $|\vec{e}_{S}|\sin \theta_D \cos \theta_S $\\
  $\vec{C}\cdot \vec{e}_{VL }$         & $|\vec{e}^{'}_{VL}| \sin \theta_D \sin \theta_S \cos\theta_{V}$\\
  $\vec{C}\cdot \vec{e}_{VR}$          & $|\vec{e}^{'}_{VR}| \sin \theta_D \sin \theta_S \cos\theta_{V} $\\
  $\vec{C}\cdot \vec{e}_{Alight}$      & $|\vec{e}_{Alight}| \sin \theta_D \sin \theta_S \sin\theta_{V} \sin\phi$\\
  $\vec{C}\cdot \vec{e}_{Aheavy\perp}$ & $|\vec{e}_{Aheavy\perp }|\sin \theta_D \sin \theta_S \sin\theta_{V} \cos \phi$\\
\hline
\end{tabular}
 \caption{Dimensionless operator coefficients expressed in the angular coordinates. The radial coordinate is $1/\LNP^2$, $\theta_I:0.. \pi$ 
  and $\phi:0..2\pi$. As discussed in Appendix~\ref{sapp:LmW}, the $\vec{e}_{VL} \times \vec{e}_{VR }$ plane was projected to a line, deviations 
  from which are measured by $\theta_V$. In general, the basis vectors $\{ e_A \}$ are not unit vectors, and their normalisation is given in 
  Table~\ref{tab:ip} and after eqn (\ref{BRBE}) for the primed vectors.
  \label{tab:angles}}
  \end{center}
\end{table}

We plot in Figure~\ref{fig:tD} the reach of $\megL$, $\meeeL$ and $\mucAlL$ as a function of $\theta_D$ for $\theta_{S} = \pi/2$, 
$\theta_{V} = \pi/4$, and $\phi = \pi/4$. This corresponds to $\vec{C} \cdot \vec{e}_{S} = 0$, so $\meee$ induced by the 
$\vec{C} \cdot \vec{e}_{D}$, $\vec{C} \cdot \vec{e}_{VR}$ and $\vec{C} \cdot \vec{e}_{VL}$, and $\muc$ probed by Al and Au. At 
$\theta_D=0,$ the dipole coefficient is only contribution to the rates. At $\theta_D = \pi/2$, 
$\vec{C} \cdot \vec{e}_{D}$ vanishes (so does $\meg$) and $\meee$ and $\muc$ are purely mediated by four-fermion operators. For 
$\theta_D> \pi/2$, $\vec{C} \cdot \vec{e}_{D}$ is negative and $\muc$ vanishes when the dipole contribution cancels the remaining 
contributions. The rate drops abruptly, indicating that the dipole contribution is relatively small and the cancellation only occurs 
in a narrow region. The valley is broader for $\meee$, since the contribution of $\vec{C} \cdot \vec{e}_{D}$ is more important, and 
the rate never vanishes because $\meee$ independently constrains each coefficient contributing to this process, so the rate only 
vanishes when all the coefficients do (see eqn ~\ref{BRmeee});
although the dipole inter-
feres with four-fermion contributions in the amplitude, the long-distance log enhancement of the dipole prevents the
vector from cancelling it in the BR.

Our  angular coordinate parametrisation defines a measure on the parameter space that assumes all the coefficients in our subspace are  ${\cal O}(1)$  once the scale $\LNP$ is fixed.
This might not be the case in some classes of models; for instance  $\vec{C} \cdot \vec{e}_{D} \gg$ four-fermion  coefficients can occur (in SUSY~\cite{Hisano:1996qq}), or the dipole could be suppressed, when the four-fermion operators are generated at tree level. 
To illustrate  complementarity  when the ``natural'' size of  $\vec{C} \cdot \vec{e}_{D}$ is significantly different from the other coefficients, 
we also plot the reach in a parametrization similar to that introduced in~\cite{deGouvea} by { defining a variable\footnote{ The original  published version of the manuscript gave a mathematically  equivalent  but analytically less practical definition.}
\beq
\kappa_D = {\rm tan}(-\theta_D)~~~.
\label{kappa}
\eeq
This non-linear transformation magnifies the regions where the dipole contribution either dominates the four-fermion interactions ($\theta\sim 0,\pi$) 
or is suppressed ($\theta\sim \pi/2$).
(A similar variable $\kappa_V = {\rm cotan} (\theta_V)$ could be defined to magnify the regions where leptonic four-fermion coefficients 
are much larger or smaller than  those with quarks.)
Notice  that $\kappa$ is a discontinuous function  of $\theta_D$
in the range $0 \to \pi$:
$\kappa_D:-\infty \to 0$  corresponds to  $\theta_D: \pi/2 \to 0$, and
$\kappa_D:0 \to \infty$ corresponds to $\theta_D:\pi \to \pi/2$.
As a result, the relative negative sign between $\tilde{C}_D$ and
$\tilde{C}_V$, which allows cancellations in the rate for $\muc$, occurs for
positive $\kappa$. (Indeed, the -1 in the argument of \ref{kappa} was chosen
for this purpose.)

The parametrisation (\ref{kappa}) allows  to write the constrainable  Lagrangian,   in terms of coefficients at  $\LNP$,  as
\bea
 {\cal L} &=& \frac{1}{\LNP^2}{\Big [}\frac{|\vec{e}_D|}{\sqrt{1 + \kappa^2}} \widetilde{O}_D
  - \frac{\kappa}{\sqrt{1 + \kappa^2}} \widetilde{O}_V {\Big ]} \label{Lag2}
\eea
where $\tilde{O}_D$ and  $\tilde{O}_V$  are respectively the dipole
and a combination of four-fermion operators  from the Lagrangian (\ref{Lag1})
---but translated from the experimental scale to $\LNP$ (see discussion below).
 The  selection of operators in  $\tilde{O}_V$   can be modified via the angles $\theta_V$ and $\phi$:
\bea
\tilde{O}_V = 
\cos \theta_V |\vec{e}_V| \tilde{ O}_{Vl}^{e\mu ee}
     +      \sin \theta_V \sin\phi |\vec{e}_{Al}|\tilde{ O}_{Alight} +
           \sin \theta_V \cos\phi |\vec{e}_{Aheavy}|\tilde{ O}_{Aheavy\perp}  
\nonumber
\eea
and  the $O(1)$ numerical factors  $|\vec{e}_X|$   encode  the rescaling of the coefficients due to  SM loop effects,  and  are given in Table  \ref{tab:ip}.
\begin{figure}[ht]
\begin{center}
 \includegraphics[width=0.4\textwidth]{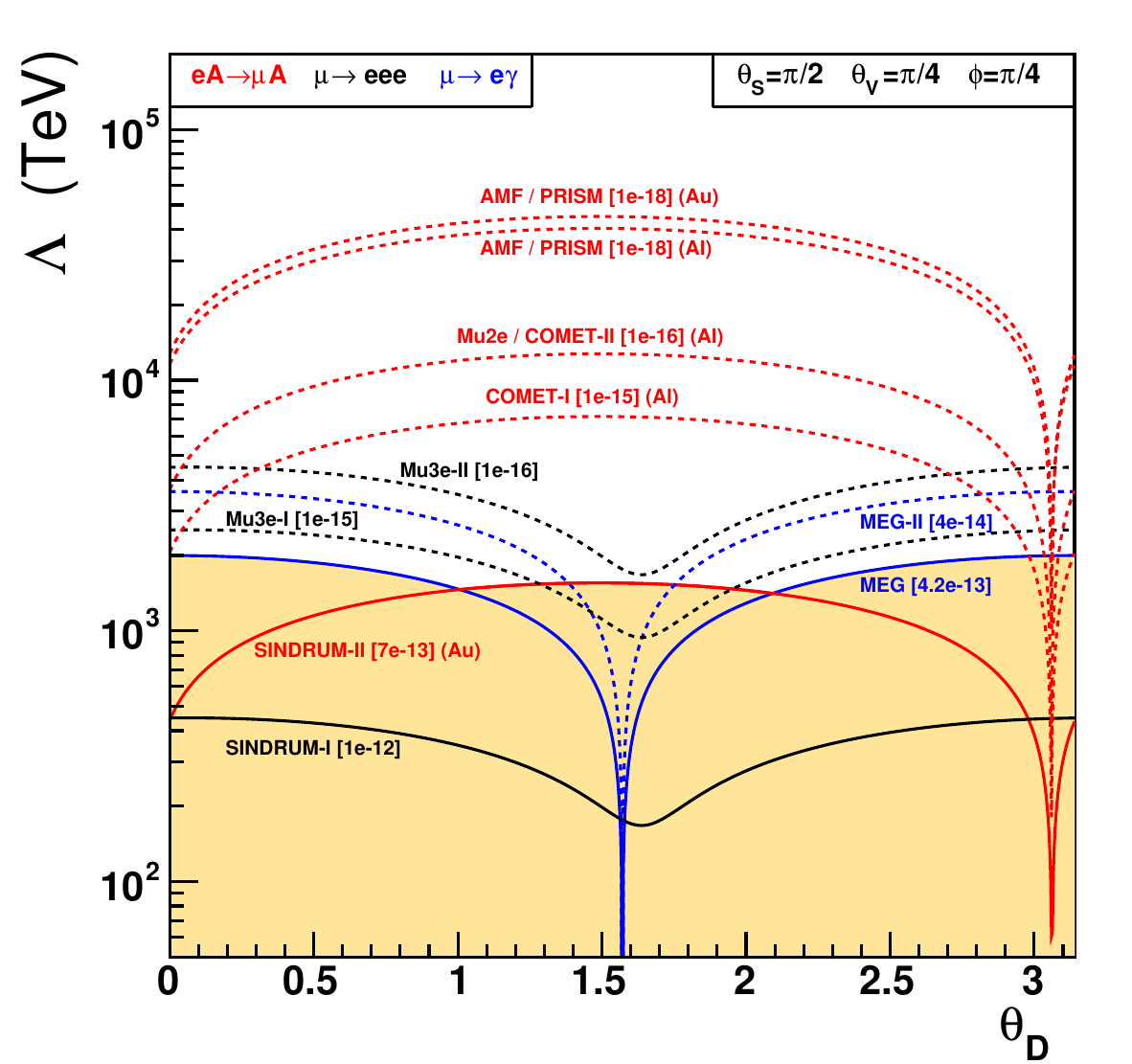}
 \includegraphics[width=0.4\textwidth]{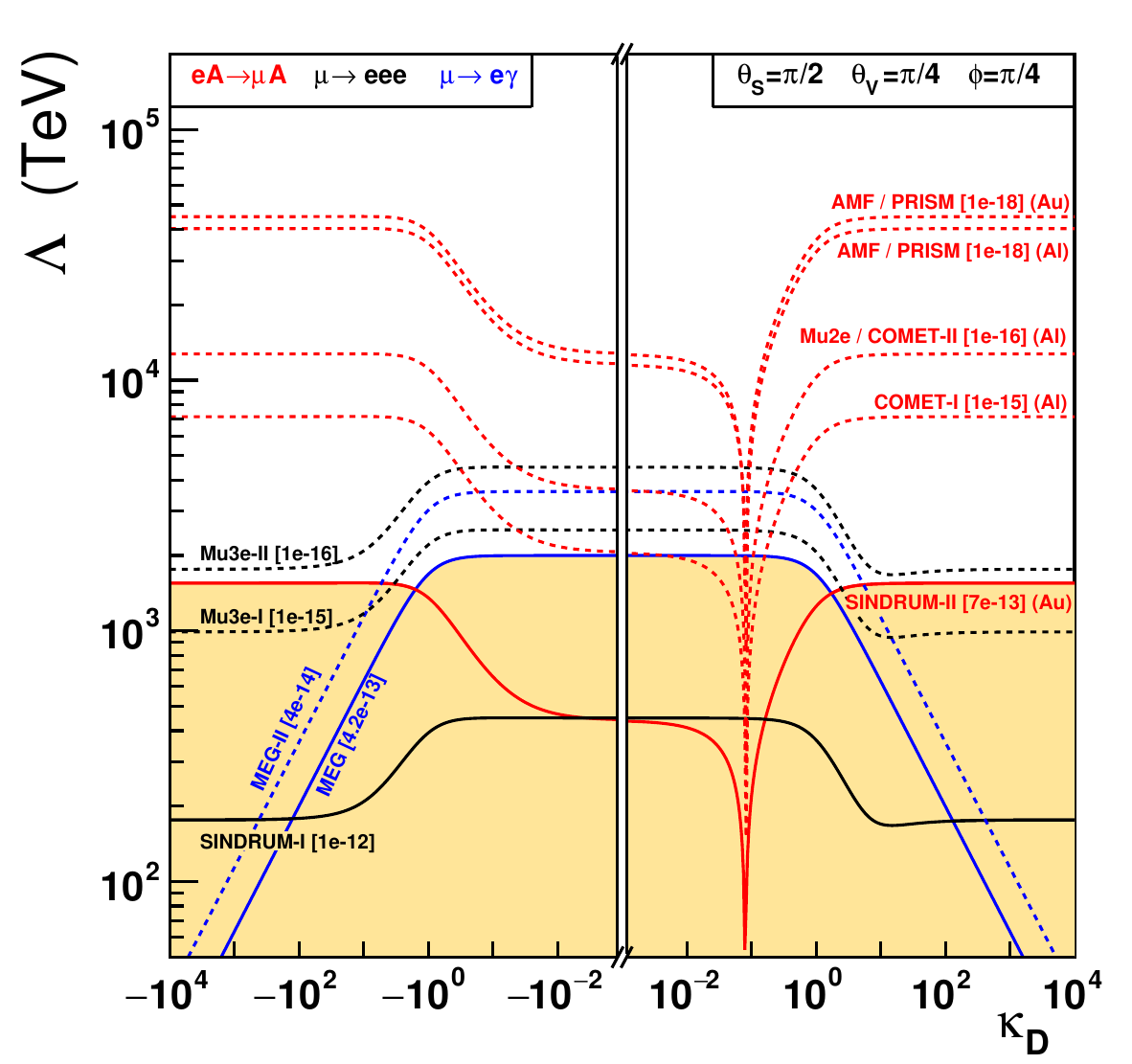}
\end{center}
\caption{Reach as a function of (left) the angle $\theta_D$, which parametrizes the relative magnitude of dipole and four-fermion coefficients, 
 and (right) the variable $\kappa_D = {\rm cotan}(\theta_D-\pi/2)$. The scale $\Lambda$ is defined in eqn (\ref{Lag1}) with the coefficients 
 normalised according to Table~\ref{tab:angles}. The solid region is currently excluded.
\label{fig:tD} }
\end{figure}

The Lagrangian of  Eq (\ref{Lag2}) is  in terms of operators and coefficients at $\LNP$,  which  are  not  identical to  those of the Lagrangian  at the experimental scale  given in Eq (\ref{Lag1}).
It is clear that the different experiments are complementary at the experimental scale $\sim m_\mu$, because they search for different processes --- however, the relevant question is  the complementarity at $\LNP$.
It is well-known that SM loops cause couplings to  grow, shrink, and mix with scale,  and it is these effects that make the difference between Eq (\ref{Lag1}) and (\ref{Lag2}). 
For instance,
the experimental bound on $\meg$, which only constrains the dipole coefficients at the experimental scale, is sensitive to a variety of other operator coefficients at
the heavy  scale $\LNP$, because  these other operators  can be  transformed to the dipole by SM loop effects. 
So  the operator $\tilde{O}_D$  is the linear combination of operators at $\LNP$   which becomes the  dipole  at low energy;  its   coefficient
$\frac{|\vec{e}_D|}{\sqrt{1 + \kappa^2}} $    is equal
to $|\vec{C} \cdot \hat{e}_D|$ appearing in Eq. (2.2),
as indicated in Eq. (2.8), so can be used to compute the Branching Ratio.
A similar mixing occurs for  four-fermion operators,  giving rise to $\tilde{O}_V$.
}

\begin{figure}[ht]
\begin{center}
 \includegraphics[width=0.32\textwidth]{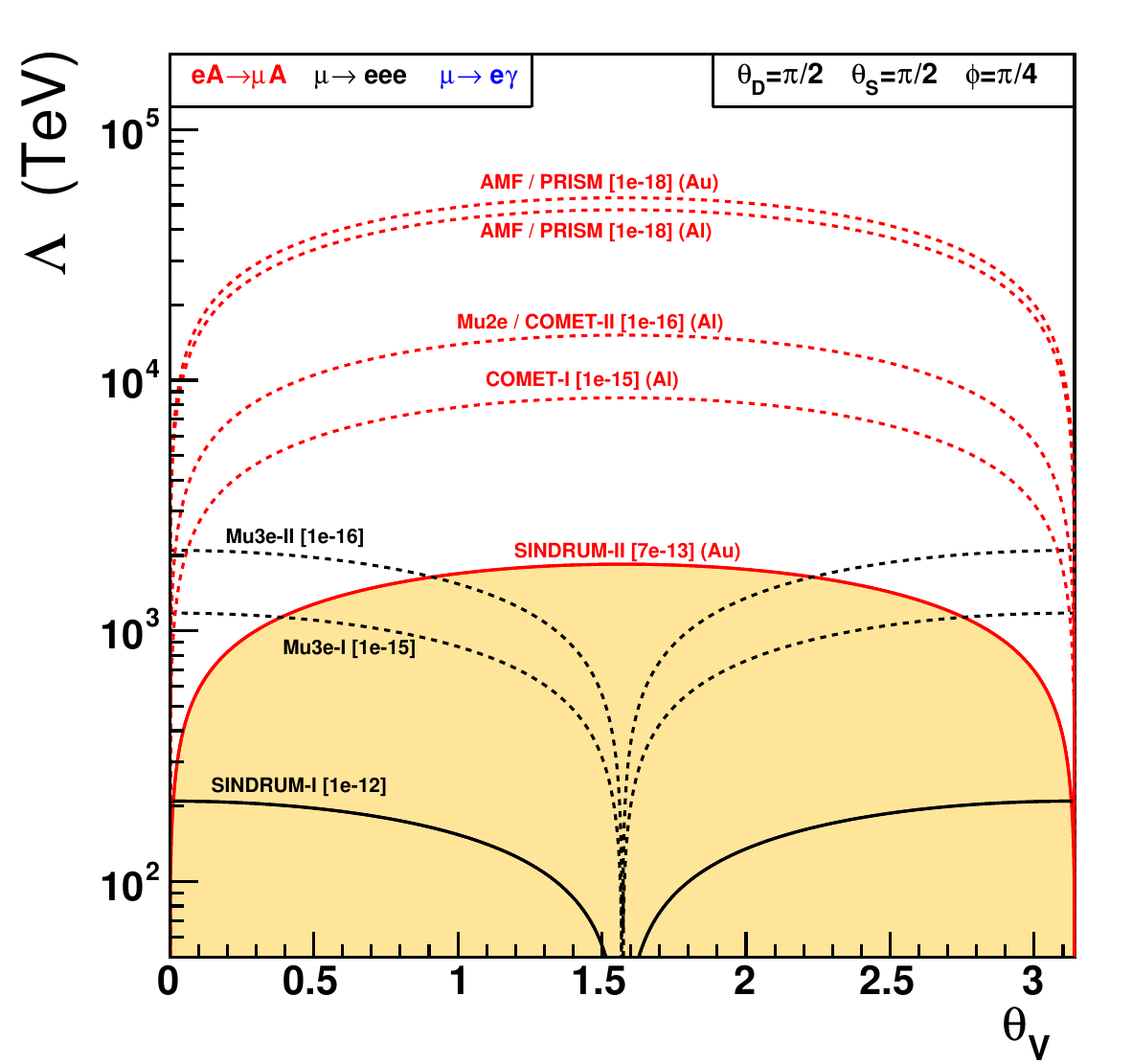}
 \includegraphics[width=0.32\textwidth]{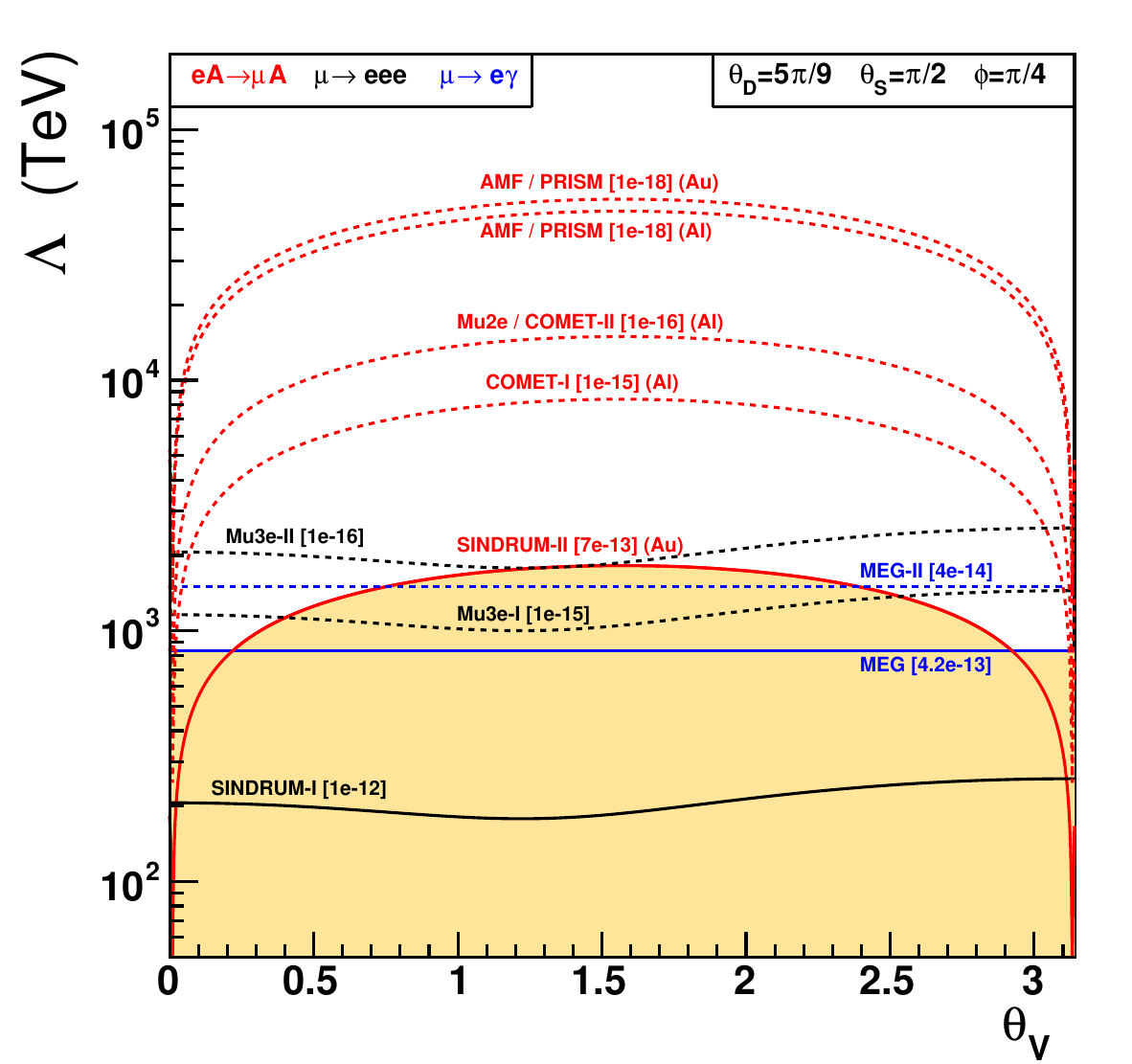}
 \includegraphics[width=0.32\textwidth]{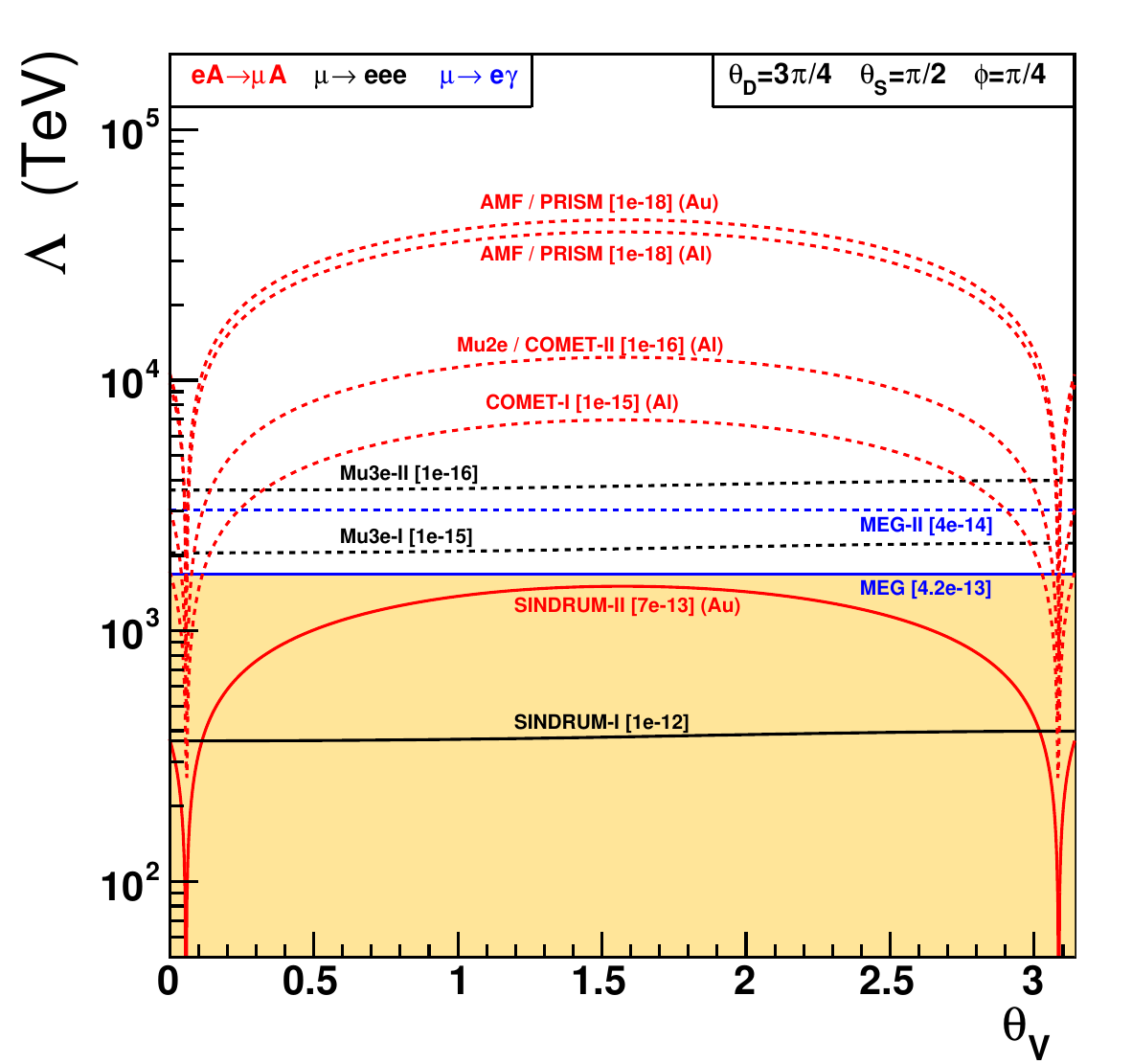}
\end{center}
\caption{Reach as a function of the angle $\theta_V$, which is effectively the angle between the $\meee$ and $\muc$
  four-fermion operators, for different contributions of the dipole operator: (left) $\theta_D=\pi/2$, (middle) $\theta_D=5\pi/9$, 
  and (right) $\theta_D=3\pi/4$. The solid region is currently excluded.
\label{fig:tV} }
\end{figure}

Figure~\ref{fig:tV} displays the reach as a function of $\theta_V$, which is effectively the angle between the $\meee$ and $\muc$ four-fermion operators. 
Results for a vanishing dipole contribution ($\theta_D=\pi/2$) shows that $\meee$ vanishes at $\theta_V= \pi/2$ and $\muc$ at $\theta_V= 0,\pi$. Adding 
a small negative dipole coefficient, $\meee$ doesn't vanish anymore since the dipole contributes independently as well as in interference with the 
four-fermion contributions, and the rate is reduced when this interference is destructive. The magnitude of the negative dipole coefficient is larger 
for $\theta_D= 3\pi/4$, exhibiting that $\muc$ vanishes when the dipole cancels the four-fermion contributions. 

\begin{figure}[ht]
\begin{center}
 \includegraphics[width=0.32\textwidth]{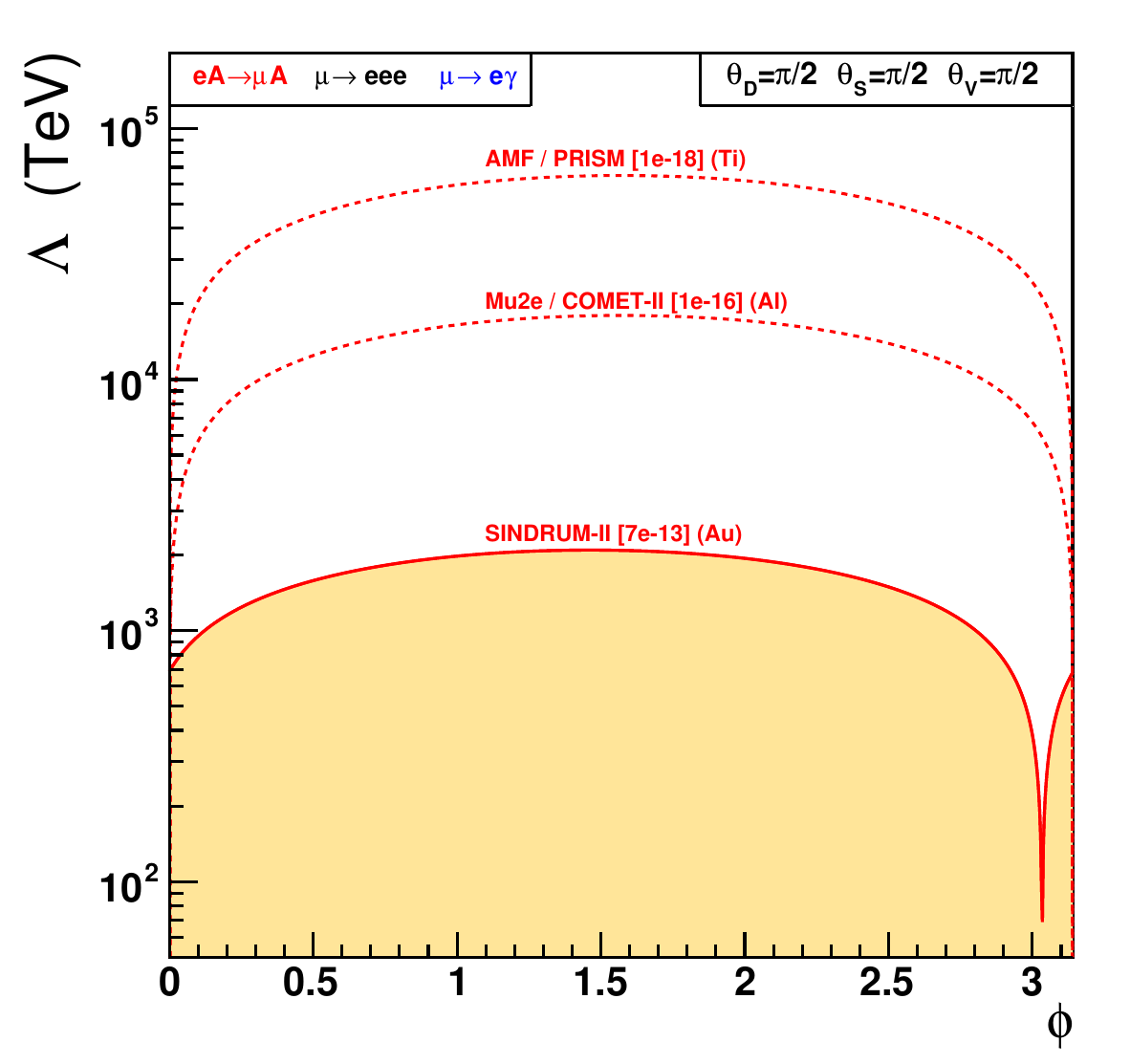}
 \includegraphics[width=0.32\textwidth]{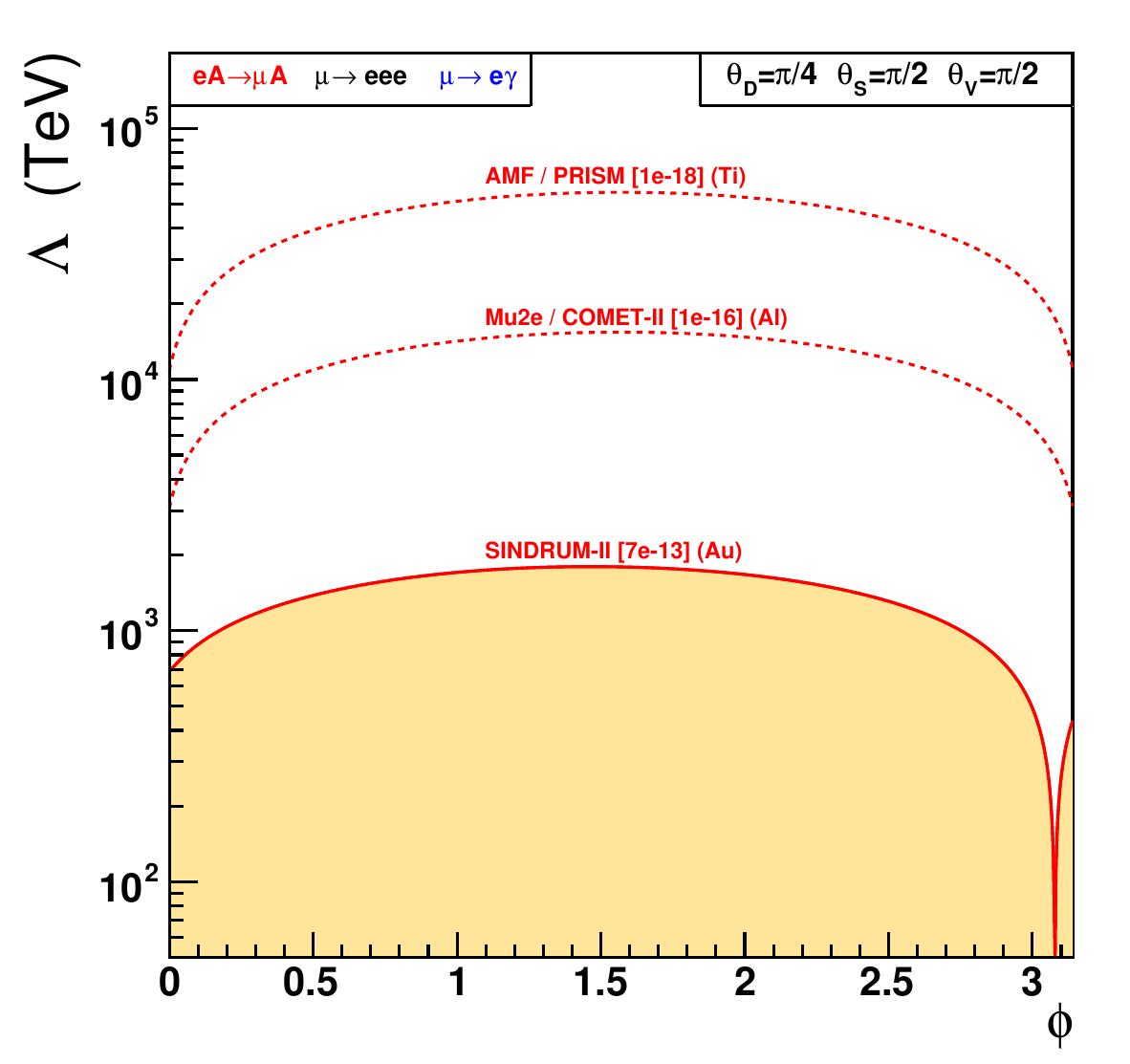}
 \includegraphics[width=0.32\textwidth]{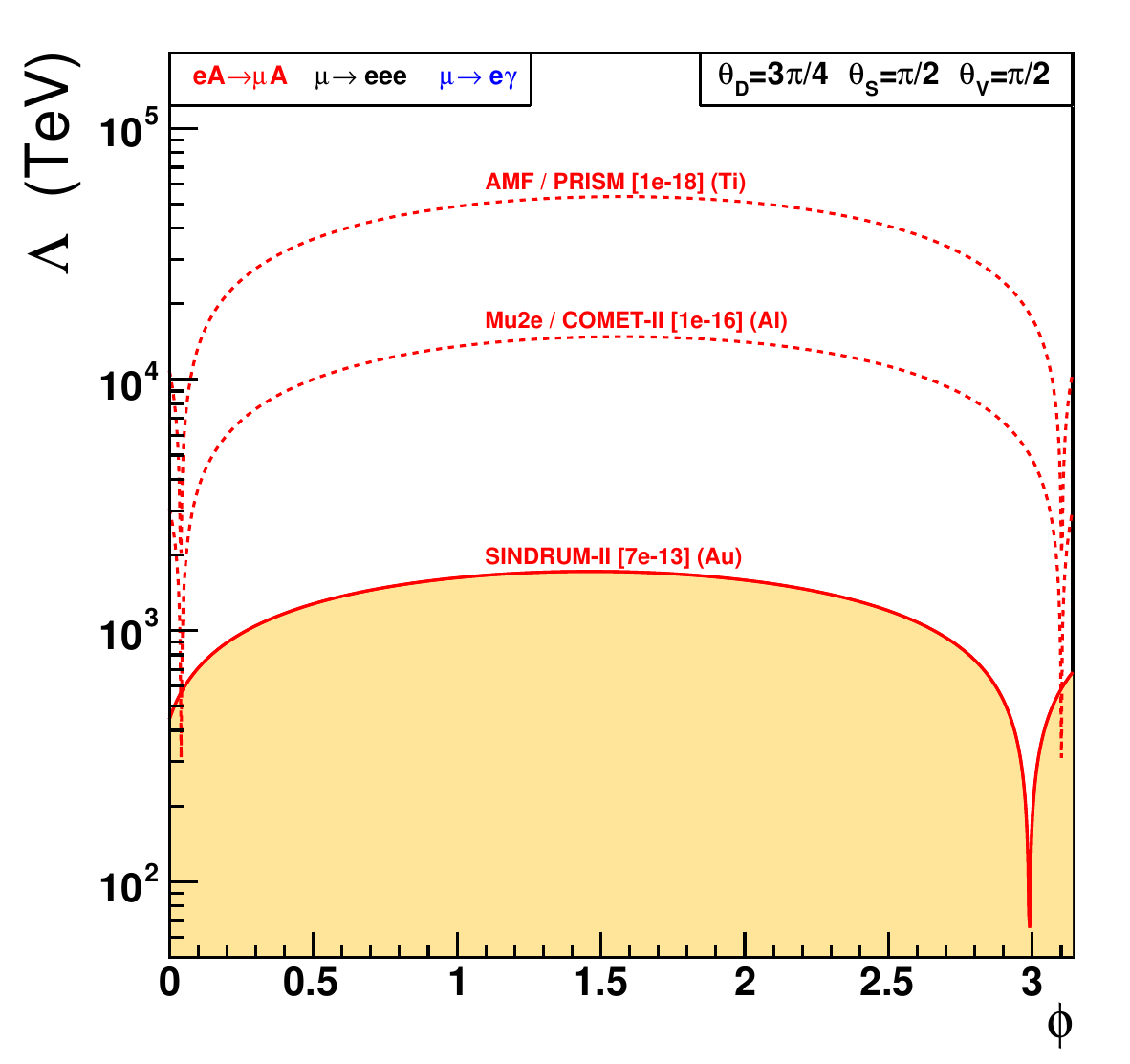}
\end{center}
\caption{Reach as a function of the angle $\phi$ for different contributions of the dipole operator: (left) $\theta_D=\pi/2$, (middle) 
$\theta_D=\pi/4$, and (right) $\theta_D=3\pi/4$. Note that $\phi$ runs from $0\to 2\pi$, although it is plotted from $0\to \pi$; the rates for 
$\phi \in(\pi \to 2\pi)$ with positive dipole are equal to those with negative dipole and $\phi \in(0 \to \pi)$. The solid region is currently 
excluded.
\label{fig:phi} }
\end{figure}

Figure~\ref{fig:phi} illustrates the complementarity of heavy and light targets for $\muc$, by plotting the conversion ratios as function of
$\vec{C} \cdot \vec{e}_{Alight} \propto \sin \phi$ and $\vec{C} \cdot \vec{e}_{Aheavy\perp}\propto \cos \phi$. Recall that 
$\vec{C} \cdot \vec{e}_{Aheavy\perp}$ parametrizes the independent information obtained with Au. This additional contribution to $\mucAuL$ 
causes the rate to vanish at a different value than that of the light targets. The dipole, which also contributes to $\muc$, was taken to 
either vanish ($\theta_D= \pi/2$), be positive ($\theta_D= 3\pi/4$) or negative ($\theta_D= \pi/4$). This illustrates the impact of 
$\vec{C} \cdot \vec{e}_{D}$ on the rate: cancellations can occur among the dipole and four-fermion contributions, as well as between 
the two independent combinations of four-fermion coefficients.

\begin{figure}[ht]
\begin{center}
 \includegraphics[width=0.45\textwidth]{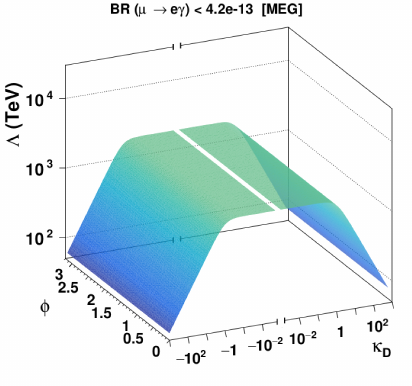}
 \includegraphics[width=0.45\textwidth]{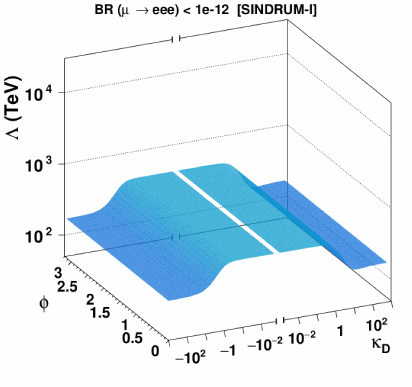}\\
 \vspace{0.5cm}
 \includegraphics[width=0.45\textwidth]{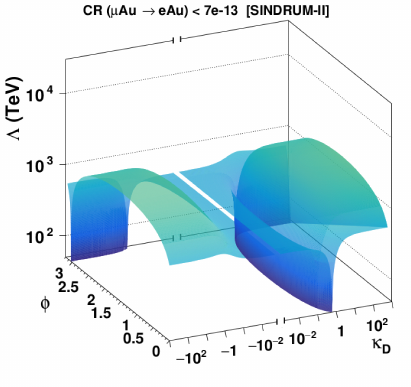}
 \includegraphics[width=0.45\textwidth]{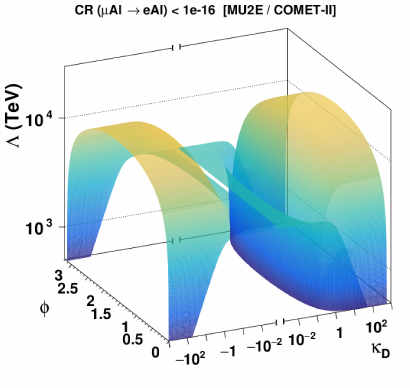}
\end{center}
\caption{Reach as a function of the angle $\phi$ and the variable $\kappa_D$ for $\theta_V=\pi/4$ and $\theta_S = \pi/2$. Note that $\phi$ 
runs from $0\to 2\pi$, although it is plotted from $0\to \pi$; the rates for $\phi \in(\pi \to 2\pi)$ with positive dipole are equal to those with 
negative dipole and $\phi \in(0 \to \pi)$.
\label{fig:phikappa} }
\end{figure}

Finally, the dependence of the sensitivity on the angle $\phi$ and the variable $\kappa_D$ is exhibited in Figure~\ref{fig:phikappa}. As expected, 
the $\meg$ and $\meee$ processes are independent of $\phi$. The shape of the conversion processes on light and heavy targets are globally similar, 
although the ridges along which the rates cancel are slightly different.

\clearpage

\section{Summary}
\label{sec:summ}

We use bottom-up EFT to calculate the reach and illustrate the complementarity of experiments searching for NP. This method is 
particularly well-suited to situations in which the number of observables is much smaller than the number of operators. It 
provides a complete parametrisation of the rates, without redundancies, and the EFT translation to $\LNP$ can be systematically 
improved. In addition, this formalism allows to explore the complementarity in a self-consistent manner at the same
scale at which the theory is defined, and ensure that experiments effectively probe different combinations of NP parameters. This
approach is generic and can be applied to many situations. In this manuscript, we use it to study CLFV in the muon sector and derive
sensitivity projections for current and future experiments.

At the experimental scale, the Lagrangian given in eqn (\ref{Lag1}) includes all and only the operators contributing at tree level
to the observables. The combinations of coefficients constrained experimentally define the operator basis for our subspace, whose dimension
is equal to the number of constraints. For $\megL$, $\meeeL$ and Spin Independent $\mucAlL$ and $\mucAuL$, this subspace is six-dimensional.
These coefficients are translated to $\Lambda_{NP}$ by solving the leading order Renormalisation Group Equations below the weak scale, and
matching them to SMEFT at tree level (see eqn (\ref{pt5})). Since the  number of constraints remains unchanged, the dimension of the 
subspace cannot grow (but it could decrease, as discussed in appendix~\ref{app:yeeks}). However, the normalisation and direction of the 
basis vectors is altered, in order to include, at $\LNP$, the contributions from all the operators  to the observables via short-distance 
effects described in the RGEs.

The ability of different experiments to probe independent operator coefficients -- our definition of complementarity -- is related to the
misalignement between vector of coefficients. While it can be measured in various ways, we observe that a judiciously selected subset of our
basis vectors remain approximately orthogonal above the weak scale, and we use various parametrisations (see table~\ref{tab:angles} or 
eqn (\ref{kappa})) to plot the experimental exclusion curves using the Branching Ratios given in eqn (\ref{BRBE}). We also  display a few 
projections to illustrate the reach and complementarity of future experiments.

An example of distinct observables probing the same New Physics is recalled in appendix~\ref{app:yeeks}: $\muc$ on various nuclei could distinguish
scalar $\mu\to e$ contact interactions on neutrons from protons, but this may not allow the distinction of LFV scalar operators involving up quarks
from those with down quarks. Improving the precision of the scalar $\bar{q}q$ expectation values in the nucleon would be required to improve
the situation.

This work is only a preliminary implementation of bottom-up EFT, relying on theoretical formalism described in~\cite{C+C}. In future work, 
we aim to implement the Renormalisation Group running of our vectors above the weak scale (it was neglected here for simplicity and the 
lack of knowledge of $\LNP$), and  match  models  onto  the ``observable subspace'' at  $\LNP$. We hope that finding robust distinctions 
among model predictions could be simplified by the reduced dimension of the subspace.

\subsection*{Acknowledgements}
SD thanks A Ibarra for a seminar invitation and discussions.

\appendix

\section{Operators, Rates and R-matrices at low energy}
\label{app:ops}

The operators contributing at the experimental scale $\sim m_\mu$ to the tree level amplitudes for $\megL$, $\meeeL$ and Spin Independent 
$\mucL$ are given in eqn (\ref{Lag1}), where the $\muc$ operators can be expanded on:
\bea
{\cal O}^{NN}_{S,R} = \overline{e} P_{R} \mu \overline{N} N
&~,~&
{\cal O}^{NN}_{V,L} = \overline{e}\gamma^{\a} P_{L} \mu \overline{N} \g_\a N
 \label{opsexpt} 
\eea
where $N\in\{p,n\}$, and the list neglects subdominant operators such as $\overline{e} P_{X} \mu FF$~\cite{DKUY}. The operator ${\cal O}_{Alight}$ is approximatively
\bea
    {\cal O}_{Alight}\propto \frac{1}{2}\left( {\cal O}^{nn}_{S,R} +
    {\cal O}^{pp}_{S,R}+
    {\cal O}^{nn}_{V,L}+{\cal O}^{pp}_{V,L}\right)~~.
\label{A2}
\eea

The Branching Ratios (BRs) for the various processes can be expressed in terms of the operator coefficients $\vec{C}$ and the matrices {\bf R}: 
\bea
BR(\megL)
 &= & \frac{v^4}{\LNP^4} \vec{C} ^\dagger \bm{R}_{\megL} \vec{C} 
 \nonumber\\
 &\Leftrightarrow & \bm{R}_{\megL}(m_\mu) = 384\pi^2  \frac{v^4}{\LNP^4} \hat{e}_{D} \hat{e}^\dagger_{D}
\label{Rmeg}
\eea
where $v\simeq m_t$ is the Higgs vev, and $\vec{C} \cdot \hat{e}_{D}={C}_{D}$. For $\meeeL$, the BRs given in the text can be written using:
\bea
 \bm{R}_{\meeeL }(m_\mu) &=& \frac{v^4}{\LNP^4}
\left[\begin{array}{cccc}
2&0 &0& 8e\\
0&1 &0& 4e\\
0&0&\frac{1}{8} &0\\
8e&4e&0&8e^2(8 \ln\frac{m_\m}{ m_e} -11 )\\
\end{array}\right]
\label{Rmeee}
\eea
where {\bf R} is given in the basis $(C^{e\mu ee}_{VL},C^{e\mu ee}_{VR},C^{e\mu ee}_{S}, C^{e\mu ee}_{D})$.

For $\muc$, we consider the conversion ratios on prototypical heavy and light targets, taken to be Au and Al\footnote{SINDRUM 
searched for $\mec$ on Titanium~\cite{Bertl:2006up}, and we will use these results in our study. However, according to~\cite{DKY}, 
Ti and Al probe the ``same'' operator coefficients within current uncertainties, so we will apply the Ti bounds along the direction in 
coefficient space corresponding to Al.}. Following Kitano, Koike and Okada (KKO)~\cite{KKO}, the Spin-Independent conversion rate 
can be written
\bea
{\rm CR}_{SI}(\mu A \to e_L A) &=&  \frac{4 m_\m^5 }{\LNP^4 \Gamma_{capt}}  
 \big|   
 {C}^{pp}_{V,L} V_A^{(p)} + {C}^{pp}_{S,R} S_A^{(p)}
+ {C}^{nn}_{V,L} V_A^{(n)} + {C}^{nn}_{S,R} S_A^{(n)} 
+ C_{D,R} {\frac{D_A}{4}} 
 \big|^2  
 \label{BRmecKKO}
 \eea
where we use the nucleus($A$)- and nucleon($N$)-dependent ``overlap integrals'' $V_A^{(N)}$, $S_A^{(N)}$, $D_A$
given by KKO~\cite{KKO}\footnote{The ``$4/\LNP^4$'' in eqn (\ref{BRmecKKO}) differs from the ``$2G_F^2$'' given in KKO's 
eqn (14), because $-4 \widetilde{C}|_{\rm here} = \tilde{g}|_{KKO}$ in the case of four-fermion operators; and $D_A$ is 
divided by 4 in the above because the dipole normalisation here is identical to KKO. In addition, the KKO overlap integrals
are in units of $m_\mu^{3/2}$, which here sits in front.}, and $\Gamma_{cap}$ is the rate for the muon to transform to a 
neutrino by capture on the nucleus. Some relevant rates are~\cite{Suzuki:1987jf}:
\beq \Gamma_{capt} = g_A^{cap} \times 10^6 {\rm sec}^{-1} = \left\{ \begin{array}{cc}
Al & 0.7054\\
Ti & 2.59\\
Au&13.07
\end{array} \right\}\times 10^6 {\rm sec}^{-1}
\label{capt}
\eeq

KKO observed that the overlap integrals were nucleus-dependent, and measurements of $\mec$ on different targets could be used to
determine the operator coefficients. Reference~\cite{DKY} explored this issue quantitatively, and showed that with current uncertainties, 
Ti and Au give independent constraints. In this work, we are interested in a slightly different question: whether the observables give 
different constraints on New Physics heavier than $m_\mu$ and $\Lambda_{QCD}$, instead of understanding if they are independent.
Unfortunately, there is ``information loss'' in matching nucleons onto quarks (see section~\ref{app:yeeks}), so we match the nucleon 
onto quark operators before constructing the {\bf R}-matrices for $\mucAlL$ and $\mucAuL$.

Nucleon operators can be matched at 2 GeV onto light quark operators (see~\cite{C+C} for a basis) as
\bea
\langle N(P_f)| \bar{q}(x) \Gamma_O q(x)|N(P_i) \rangle \simeq
G^{N,q}_O \langle N (P_f)| \bar{N}(x) \Gamma_O N(x)|N (P_i)\rangle=G^{N,q}_O\overline{u_N}(P_f) \Gamma_O u_N(P_i) e^{-i(P_f-P_i)x} \nonumber ~~~.
\eea
We also include the two-step matching of scalar $b$ and $c$ operators $(\bar{e} P_R \mu)(\bar{Q} Q)$~\cite{SVZ,CKOT}, first onto the gluon operator $(\bar{e} P_R \mu)GG$, then onto nucleons. As a result, the nucleon and quark coefficients are related as
\bea
{C}^{NN}_{O,X} = \sum_q G_O^{Nq} C_{O,X}^{qq}~~~,
\label{text}
\eea
where $O\in\{S,V\}$ and the relevant $G$s are given in Table~\ref{tab:G}. One can then define quark ``overlap integrals'' for target $A$ as 
\bea
I^q_{A,S}& =& G_{S}^{pq} S_A^p + G_{S}^{nq} S_A^n \nonumber\\
 I^q_{A,V}& = &G_{V}^{pq} V_A^p + G_{V}^{nq} V_A^n
\label{quarkOI}
\eea
where in this work we use the EFT results~\cite{Hoferichter:2015dsa} for $G_S^{N,q}$, which differ by $\sim$50\% from lattice 
results~\cite{Lellouch,LLF}. Assembling the quark overlap integrals for target $A$ into a ``target vector'' $\vec{u}_A$: 
\bea
\vec{u}_{A} &=& (I^b_{A,S}, I^c_{A,S},I^s_{A,S},I^u_{A,S},I^d_{A,S},
I^u_{A,V},I^d_{A,V}) ~~,
\label{uA}
\eea
allows to write the Conversion Ratio as 
\bea
  {\rm CR}_{SI}(\mu A \to e_L A)
   &=& \frac{6144 \pi^3v^4}{2.197 g^{cap}_A \LNP^4}  
 \left|   \vec{u}_A \cdot \vec{C} 
+ C_{D} \frac{D_A}{4} 
\right|^2
\nonumber\\
&\equiv& \vec{C}^\dagger \bm{R}_{\mucL} 
\vec{C} 
\label{preR2G}
\eea
where  $g^{cap}_A$ is defined  in Eq. (\ref{capt})  and $D_A$ is the KKO overlap integrals for the dipole~\cite{KKO}. So the {\bf R}-matrix at a scale of 2 GeV is
\bea
\bm{R}_{\mucL} &=& \frac{6144 \pi^3v^4}{2.197 g^{cap}_A \LNP^4}
  {\Big (}\vec{u}_{A} + \frac{D_A}{4} \hat{e}_D{\Big )} \otimes
 {\Big (} \vec{u}_{A} 
+ \frac{D_A}{4} \hat{e}_D{\Big )}^\dagger
\label{Rmuc}
\eea

This translation of nucleon to quark operators neglects higher order QED and ``strong interaction'' effects between the 
experimental scale $m_\mu$ and 2 GeV: the QED running is small for the considered operators, and we did not include recent 
$\chi$PT results ~\cite{VCetal}.

The prospects of distinguishing coefficients by using different targets depend on the misalignment between the target 
vectors, which can be quantified as an angle\cite{DKY}:
\beq
 \vec{u}_{A_1} \cdot \vec{u}_{A_2} = |\vec{u}_{A_1}| | \vec{u}_{A_2}| \cos \theta~~.
\label{thetadef}
\eeq
In the quark operator basis at 2 GeV, this angle is given for various potential targets in Table~\ref{tab:misalignment2G}. 
The angles are smaller than the misalignmeent angles in the nucleon operator basis\cite{DKY} (see Figures~\ref{fig:thetan} and ~\ref{fig:thetaq}), 
because the scalar quark overlap integrals are larger than the vector integrals, and comparable for $u$ and $d$ quarks. In 
addition, there is currently a large discrepancy, $\sim 50$\%, between lattice and EFT determinations of the scalar 
overlap integrals. We assume that this theoretical discrepancy can be solved, so that $\mucL$ on Au and Al give 
independent information.

\begin{table}[ht]
$ 
\begin{array}{|c|cccccc|}
\hline
&{\rm Li} & {\rm Al}& {\rm Ti} &{\rm Cu} & {\rm Au}& {\rm Pb} \\
\hline
{\rm Li} &- & 0.545 &  0.714 & 1.06&  5.57 & 6.69 \\
{\rm Al} &&- & 0.433 & 0.760 & 5.34 & 6.46 \\
{\rm Ti} &&&- &0.356 & 4.94 &6.06\\
{\rm Cu} &&&&- &4.59 & 5.71 \\
{\rm Au} &&&&&- &1.12 \\
\hline
\end{array}
$
\caption{Misalignment angles (in degrees), between target vectors expressed in the quark operator basis.
\label{tab:misalignment2G}
}
\end{table}

Eqn (\ref{Rmuc}) gives the {\bf R}-matrices for Al and Au, which probe directions in quark coefficient space that are misaligned
by $\sim$ 5 degrees, at a scale of 2 GeV. In the plane spanned by $\vec{u}_{Au}$ and $\vec{u}_{Al}$, the orthogonal combinations 
used in eqn (\ref{Lag1}), can be obtained by writing the target vector $\vec{u}_{Au} $ for Au as:
\bea
\vec{u}_{Au}&= &|\vec{u}_{Au}| (\cos\theta_{\rm AuAl} \hat{u}_{Al} + \sin\theta_{\rm AuAl}
\hat{u}_{\perp}) \nonumber\\
\Rightarrow \hat{u}_{\perp}& =& \frac{ \hat{u}_{Au} -\cos\theta_{AuAl} \hat{u}_{Al} }{\sin\theta_{AuAl}} 
~~~~~~~~~~~~~~~~~~~~~~~~
\label{thetaAlAu}
\eea
where 
$\theta_{AuAl}$ is given in Table~\ref{tab:misalignment2G}, and $\vec{u}_{\perp} \approx .56 \hat{e}^{uu}_{V} + .8 \hat{e}^{dd}_V$ is the direction 
in coefficient space corresponding to the operator ${\cal O}_{Aheavy\perp}$ of eqn (\ref{Lag1}). As a result, from Eq.s  (\ref{A2}),(\ref{thetaAlAu}) and
(\ref{quarkOI}), one approximatively obtains
  \bea
      {\cal O}_{Alight} &\approx & \frac{1}{\sqrt{2}}
      \left( O_S^{uu} + O_S^{dd}\right) + \frac{1}{16} \left( O_V^{uu} + O_V^{dd}\right)
\nonumber\\
    {\cal O}_{Aheavy\perp} &\approx& \frac{1}{5} \left(3 O_V^{uu} + 4O_V^{dd}\right)~~.
    \label{A14}
    \eea


\section{Are scalar quark currents indistinguishable?}
\label{app:yeeks}

This Appendix is 
focused on the information loss that occurs in matching
nucleons to quarks at 2 GeV. In the data, LFV scalar interactions
with neutrons might be distinguishable from those on protons~\cite{DKY}. 
But current theoretical translations of the nucleon results to quarks erase any distinction between LFV interactions
with scalar $u$ or $d$ currents (the subdominant $s,c,b,t$ quarks are neglected in this section).

The Spin-Independent Conversion Rate (CR) is given in eqn (\ref{BRmecKKO})
in terms of operator coefficients on nucleons. This result is at
``Leading Order'' in the low energy theory, and does not include
the next order in $\chi$PT (parametrically $\sim 10$\%, two-nucleon 
effects, pion exchange...) or in the nuclear matrix element.
Such effects have been calculated for WIMP scattering~\cite{Hetal},
and some partial results for $\muc$ have been obtained~\cite{Crivellin:2014cta,VCetal,Dekens}.
Improvements of the theoretical calculation of $\Gamma(\muc)$ 
could change the form of the conversion ratio (this occurs in~\cite{VCetal}),
or reduce the uncertainties on its parameters, thereby resolving
the issue discussed here.

To set the stage, recall from KKO~\cite{KKO} that the overlap integrals for scalar and vector densities of neutrons and protons differ by 
less than a factor of three. In particular, Al approximately probes
\beq
C^{pp}_{S,R} + C^{pp}_{V,L} + C^{nn}_{S,R} + C^{nn}_{V,L}~~, 
\label{combo1}
\eeq
(neglecting the dipole which could be constrained/measured elsewhere). In addition, it was pointed out in~\cite{DKY} that measuring $\muc$
on another light target with different numbers of $n$ and $p$ would allow the measurement of
\beq
C^{pp}_{S,R} + C^{pp}_{V,L} - C^{nn}_{S,R} - C^{nn}_{V,L}~~. 
\label{combo2}
\eeq
Then, as noted by KKO, vector overlap integrals dominate over the scalars in heavy targets, so comparing Al to Au could allow 
to determine
\beq
C^{pp}_{S,R} - C^{pp}_{V,L} + C^{nn}_{S,R} - C^{nn}_{V,L}~~. 
\label{combo3}
\eeq
However, heavy targets also have more neutrons,
so the $n-p$ measurement from comparing two light targets
is required to extract the $S-V$
from heavy targets. The one remaining combination of coefficients
(an isospin-violating $V$-$S$ difference) has little impact on the CR given 
in eqn (\ref{BRmecKKO}), and cannot be extracted with current theoretical uncertainties~\cite{DKY}.

\begin{table}[th]
\begin{center}
\begin{tabular}{|l|l|l|}
\hline
$G^{p,u}_V = G^{n,d}_V = 2$ &$ G^{p,d}_V = G^{n,u}_V = 1$&$ 
G^{p,s}_V = G^{n,s}_V = 0$ 
\\
\hline
$G^{p,u}_S = \frac{m_p}{m_u} 0.021(2) =9.0$& $G^{p,d}_S = \frac{m_p}{m_d} 0.041(3) = 8.2$&  \\ 
$[G^{p,u}_S = \frac{m_p}{m_u} 0.0139(13)(12) =5.9 ]$ &
$[G^{p,d}_S = \frac{m_p}{m_d} 0.0253(28)(24) =5.0]$
 &$ G^{p,s}_S =  \frac{m_N}{m_s} 0.043(11) = 0.42$ 
\\
 $G^{n,u}_S = \frac{m_n}{m_u} 0.019(2)= 8.1$
&$G^{n,d}_S = \frac{m_n}{m_d} 0.045(3)= 9.0$& \\
$[G^{n,u}_S = \frac{m_n}{m_u} 0.0116(13)(11)=5.0]$ &
$[G^{n,d}_S = \frac{m_n}{m_d} 0.0302(3)= 6.0]$& 
$G^{n,s}_S =  \frac{m_N}{m_s} 0.043(11) = 0.42$\\
&&$G^{N,c}_S =0.0543(= \frac{2m_N}{27m_c}f_{GG}^N) $\\
&&$G^{N,b}_S =0.0158(= \frac{2m_N}{27m_b}f_{GG}^N)$\\
\hline
\end{tabular}
\caption{\label{tab:G}
  Expectation values of vector and scalar quark currents in the nucleon;
  the scalar values are sometimes 
  called $f^N_q$~\cite{LLF},  and  for light quarks are related to 
  $\sigma_{\pi N}$~\cite{Hoferichter:2015dsa}, where 
  $G_S^{N,q} = \frac{m_N}{m_q} f^N_q = \sigma_{\pi N}/m_q$+ isospin corrections.
  The first scalar $G_S$ induced by $u$ and $d$  quarks 
  were obtained via 
 dispersive techniques~ \cite{Hoferichter:2015dsa,Crivellin:2013ipa,RuizdeElvira:2017stg} 
 and EFT methods~\cite{ Alarcon:2011zs} (Some lattice calculations give similar results ~\cite{Gupta:2021ahb}).
 The second $G^{N,q}_S$ [in brackets]
   are the lattice results of BMW~\cite{Lellouch}, and an
  average of lattice results~~\cite{Junnarkar:2013ac} is used 
  for the strange quark.
  The heavy quark scalar $G_S$s are from the lattice~\cite{LLF},
  and in parentheses from ~\cite{SVZ}.
  In all 
  cases, the $\overline{\rm MS}$ quark masses at $\mu = 2$~GeV
  are taken as $m_u = 2.2$~MeV, $m_d = 4.7$~MeV, 
  and $m_s = 96$~MeV~~\cite{Agashe:2014kda}.
  The running heavy quark masses are~\cite{PDB} $m_c(m_c) = 1.27$~GeV, 
 $m_b(m_b) = 4.18$~GeV. The nucleon masses are $m_p = 938 ~{\rm MeV}$ and $m_n = 939.6 ~{\rm MeV}$.
}
\end{center}
\end{table}

The operator coefficients on nucleons can be transformed to coefficients on quarks according to eqn (\ref{text}),
using the matrix $G$ given in Table~\ref{tab:G}\footnote{This transition to quarks was not included in~\cite{DKY} due to the 
discrepancy among theoretical determinations of the $G_S^{N,q}$.}. Focusing on the first generation valence quarks, 
this can be written as:
\beq
\left(\begin{array}{c}
C^{pp}_{V,L}\\
C^{nn}_{V,L}\\
C^{pp}_{S,R}\\
C^{nn}_{S,R}\end{array}
\right)
=
\left[\begin{array}{cccc}
2&1&0&0\\
1&2&0&0\\
0&0&G_S^{pu}&G_S^{pd}\\
0&0&G_S^{nu}&G_S^{nd}
\end{array}
\right]
\left(\begin{array}{c}
C^{uu}_{V,L}\\
C^{dd}_{V,L}\\
C^{uu}_{S,R}\\
C^{dd}_{S,R}\end{array}
\right)
\label{Gm}
\eeq
where the vector coefficients exhibit the expected dominance of $u$ quarks in the proton, and $d$ quarks in the neutron: the quark 
vector coefficients can be calculated from the nucleon vector nucleon coefficients, and vice versa. In the case of the scalar coefficients, 
this is almost not the case; for both lattice and EFT determinations of the $G_S^{N,q}$, the determinant of the scalar submatrix in 
eqn (\ref{Gm}) is small compared to the product of two $G_S^{N,q}$ ($G_S^{pu}G_S^{nd} -G_S^{pd}G_S^{nu}$ $\sim 9^2 -8^2 (6^2 - 5^2)$),
causing large uncertainties when it is inverted to solve for quark coefficients as a function of nucleon coefficients. In addition, 
$G_S^{N,q}\gg G_V^{N,q}$ increases the sensitivity to scalar coefficients, and reduces the relative contribution of the vector 
coefficients to about the magnitude of the scalar uncertainties. 

\begin{figure}[ht]
\begin{center}
 \includegraphics[scale=0.5]{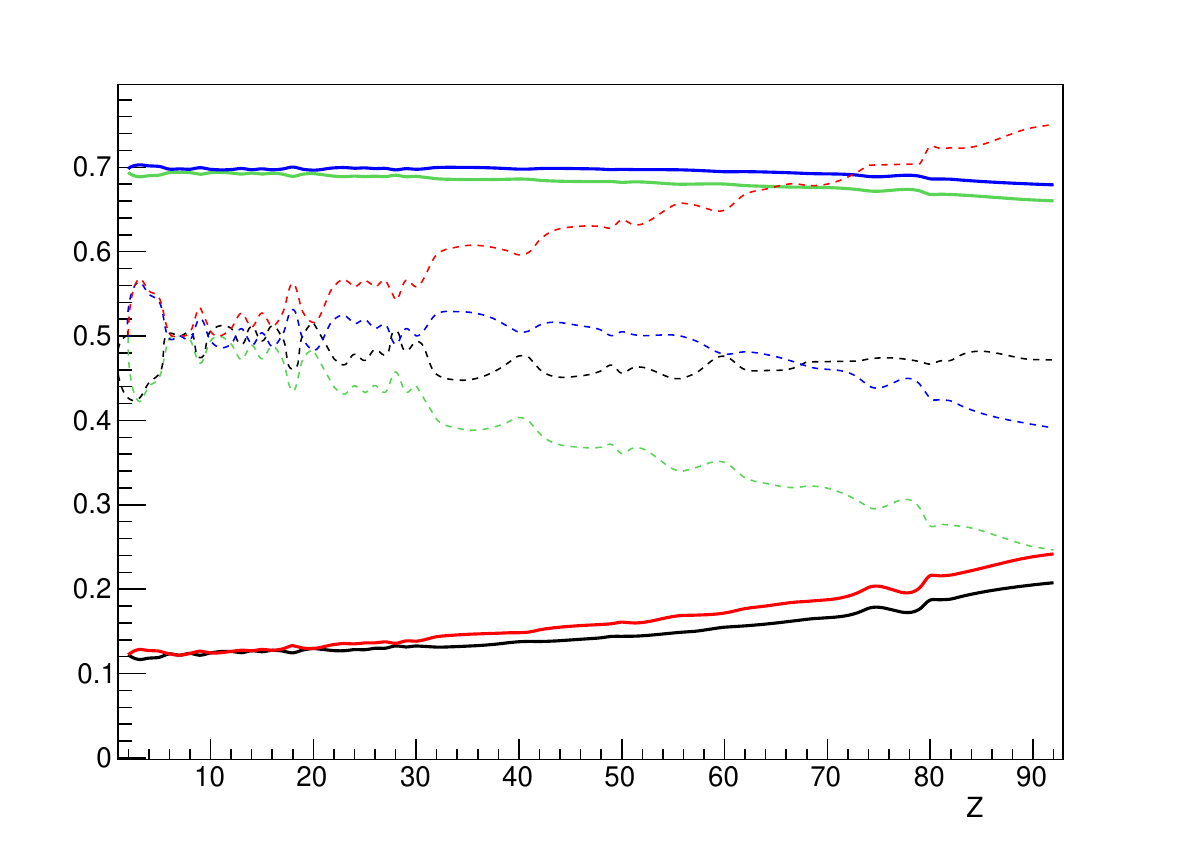}  
\end{center}
\caption{Normalised quark (solid line) and nucleon (dotted line) overlap integrals as a function of the target charge $Z$. From the top-down 
 at $Z=$ 90: vector $n$ (red, dotted), scalar $d$ (blue) and $u$ (green), vector $p$ (black,dotted), scalar $n$ (blue,dotted) and 
 $p$ (green, dotted), and vector $d$ (red) and $u$ (black).
The  nucleon  overlap  integrals are  defined in Eq. (\ref{BRmecKKO}), and those for quarks in  Eq. (\ref{quarkOI}); an overlap integral for nucleons  (or quarks) in target $A$ is  normalised by dividing by the 
 the square root of the  sum of the squares of all the nucleon (or quark) integrals for that target.
\label{fig:OI} }
\end{figure}

The effects of transforming from nucleon operators to quark operators can be seen in Figure~\ref{fig:OI}, where are plotted the quark and 
nucleon overlap integrals. This illustrates the difficulty to distinguish scalar $u$ vs $d$, and also that the variation with $Z$ is 
reduced for the quarks compared to nucleons.

\begin{figure}[ht]
\begin{center}
\epsfig{file=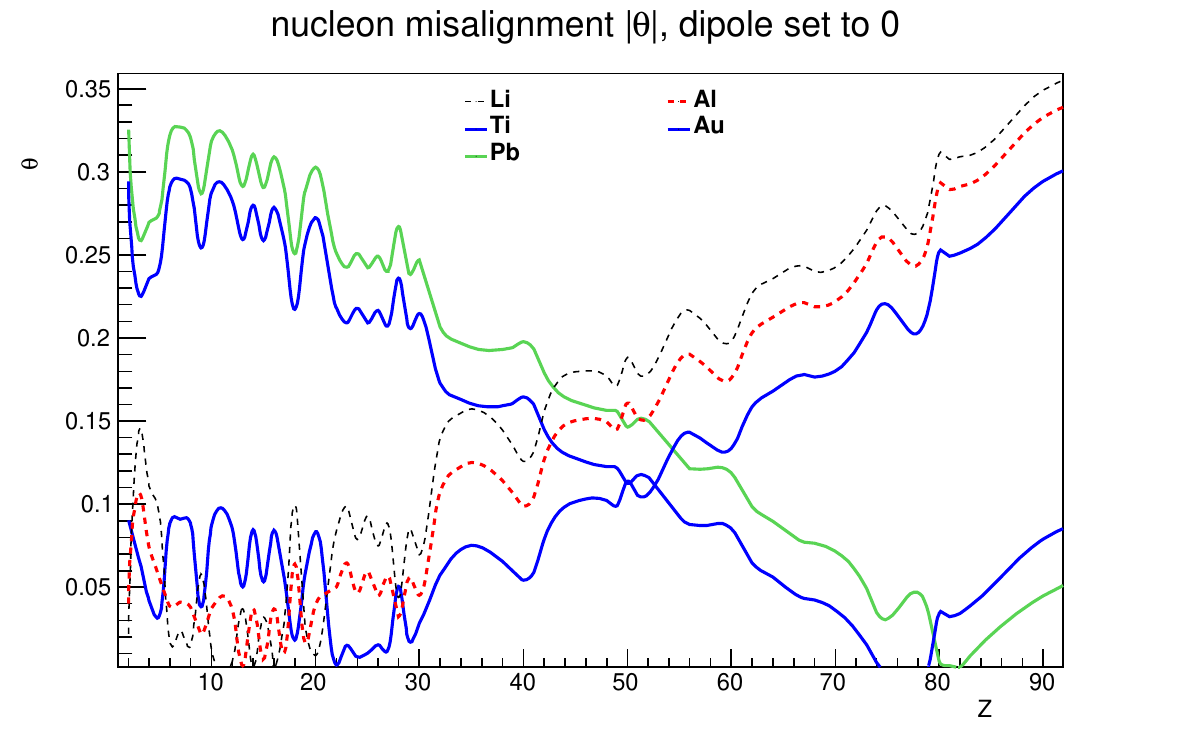,height=6cm,width=8.5cm}
\end{center}
\caption{The misalignement angle $\theta$ (defined in Eq. (\ref{thetadef})  between nucleon  target vectors.  The decreasing lines are lead (green,upper) and Au (blue); the rising lines are titanium 
 (blue) and Al (red,dashed, middle) and lithium (black,dashed, upper).
 \label{fig:thetan} }
\end{figure}

\begin{figure}[ht]
\begin{center}
\epsfig{file=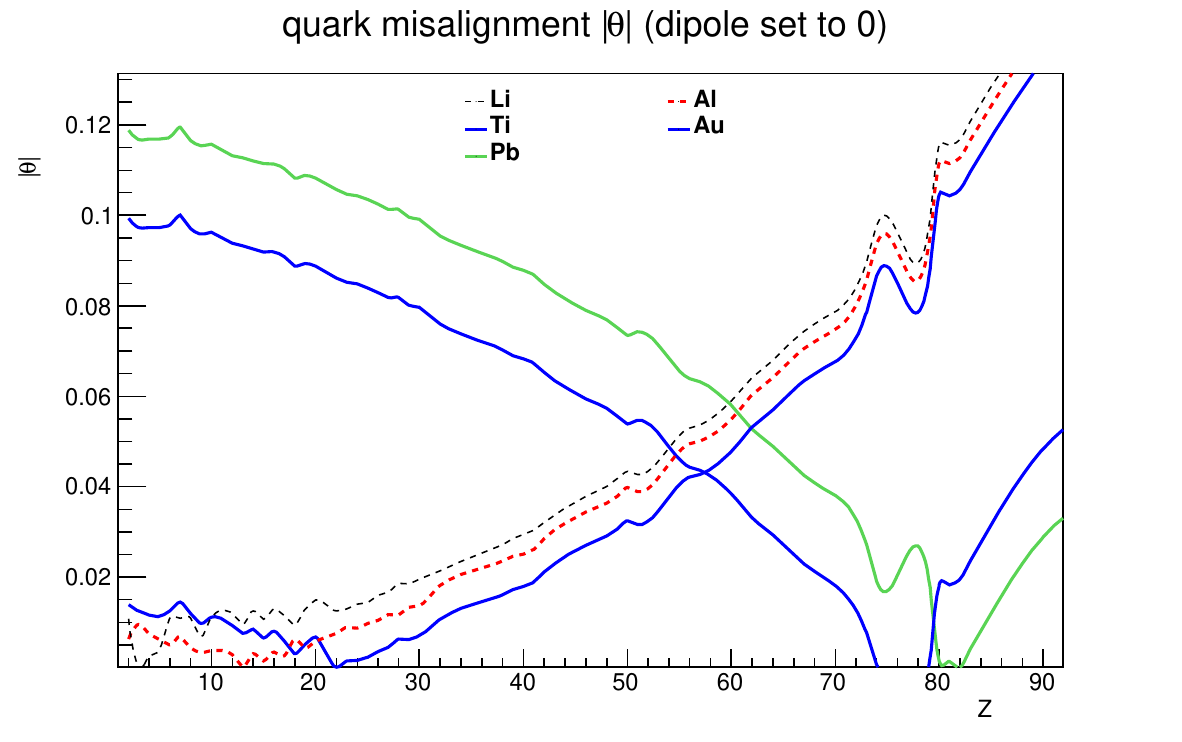,height=6cm,width=8.5cm}
\end{center}
\caption{The misalignement angle $\theta$ (defined in Eq. (\ref{thetadef}) between quark target vectors; notice the reduced vertical scale and
 smoother lines as compared to the nucleon plot. The decreasing lines are lead (green,upper) and Au (blue); the rising lines are titanium 
 (blue) and Al (red,dashed, middle) and lithium (black,dashed, upper).
 \label{fig:thetaq} }
\end{figure}

The ability of different targets to distinguish among operator coefficients can be quantified as the angle between the directions they 
probe in coefficient space~\cite{DKY}, where the direction is given by the overlap integrals for the target. Figures~\ref{fig:thetan} and ~\ref{fig:thetaq} plots 
this misalignment angle as a function of $Z$ for various past and future targets. The plot on the left is for quark coefficients, and on 
the right is for nucleon coefficients (notice the difference in the vertical scale). Reference~\cite{DKY} estimated that a misalignment 
angle $\gsim 0.1$ was required to overcome theoretical uncertainties in the nuclear calculation, and obtain independent measurement of
distinct nucleon operator coefficients. Even if one neglects the theoretical uncertainty in the $G_O^{N,q}$s, applying this $\gsim 0.1$
rule for quark coefficients suggests that the theoretical accuracy needs to be improved.

In summary, upcoming experiments could hope to bound or measure the three combinations of nucleon coefficients given in eqns
(\ref{combo1}-\ref{combo3}). However, due to the almost-vanishing scalar determinant in eqn (\ref{Gm}), theoretical progress in calculating 
the $G^{Nq}_S$ is required, for these to constrain three independent combinations of quark operator coefficients.

\section{Including the penguin}
\label{app:ping}

This Appendix discusses the orthogonality of the basis used for the experimentally accessible subspace. We need an orthogonal basis to 
illustrate complementarity in polar coordinates; orthogonality would not be required to evaluate complementarity via eqn (\ref{TrR}),
nor for exploring the model predictions for coefficients in the subspace. In the following, the term ``penguins'' refer to the operators 
of eqn (\ref{pingouinSMEFT}).

The basis vectors at $\LNP \gsim m_W$ are {\bf G}$^*(m_W,m_\mu) \hat{e}_{A}$; we do not give explicit expressions because the 
translation from the observable-motivated basis to an arbitrary other choice is a technicality more suitable to computers. Expressions 
for various $\vec{C}^\dagger \bm{ G}^* (m_W,m_\mu) \hat{e}_{A}$, appropriate for calculating LFV BRs in terms of coefficients at the
weak scale, can be found in~\cite{C+C}. The norms and inner products among some basis vectors are given in Table~\ref{tab:ip}, which 
shows an overlap of $ 10 \to 20 $ degrees between the vector operators ($e_\perp$ has significant quark-vector components).
The shrinking of the quark target-vectors is largely due to the shrinking scalar coefficients.

The overlap among the coefficients of $(\bar{e}\g^\a P_L \mu)(\bar{f}\g_\a f)$ operators arises because the $\mu\to e $ penguin operators
of the SM EFT ~\cite{polonais}:
\beq
 \d{\cal L}_{penguin}=
 \frac{ C_{HL1}}{\LNP^2}(\overline{\ell}_e \gamma^\a \ell_\mu )(H^\dagger \stackrel{ \leftrightarrow}{ D_\a } H) + \frac{C_{HL3}}{\LNP^2}
   (\overline{\ell}_e \tau^a \gamma^\a \ell_\mu )(H^\dagger \stackrel{ \leftrightarrow}{ D^a_\a } H)
   \label{pingouinSMEFT}
\eeq
generate a flavour-changing vertex $\frac{gv^2}{c_W \LNP^2 }(C^{e \mu}_{HL1} + C^{e \mu}_{HL3})
\bar{e}\Zslash P_L \mu $, so that $Z$ exchange gives four-fermion operators in the low energy EFT:
\beq
 \d {\cal L} = \sum_{f,X}\frac{g^f_X(C^{e \mu}_{HL1} + C^{e \mu}_{HL3})}{\LNP^2}
 (\overline{e} \gamma^\a P_L \mu )(\overline{f} \gamma^\a P_X f)  
\eeq
where $f$ is any light chiral fermion, we used the SM $Z$ interaction $ \frac{g}{2c_W}(g^{f}_{L} P_L + g^{f}_{R} P_R)$ and 
$c_W=\cos\theta_W$. As a result, at the weak scale, the basis vectors $\vec{e}_{VL} $ and $\vec{e}_{VR} $ have components $ g^e_{L}$ 
and $g^e_{R}$ in the two penguin directions, respectively, giving a tree-level contribution to $\vec{e}_{VL} \cdot \vec{e}_{VR}$ 
of order $ 2g^e_{L}g^e_{R} = 4 s_W^2(-1+2 s_W^2)$.

A more orthogonal basis could be obtained by removing the penguin contribution from the low energy $(\bar{e}\g^\a P_L \mu)(\bar{f}\g_\a f)$ 
operators, and adding $Z\to \mu e_L$ as an observable at $m_W$. This is analogous to what was already done for the dipole, removing it
from the combination of operators contributing to $\muc$, and including $\meg$ as an observable. However, this adds a dimension to the 
subspace, and tangles the intuitive link between basis vectors and observables. It is pursued in section~\ref{sapp:LmWps}. Section~\ref{sapp:LmW} 
describes a simpler approach used to make the plots in the body of the paper.

\begin{table}[ht]
 $
\begin{array}{|c|cccccc|}
\hline
&\vec{e}_{Alight}&\vec{e}_{Aheavy\perp}&\vec{e}_{VL} &\vec{e}_{VR}& \vec{e}_{S}&\vec{e}_{D} \\
\hline
\vec{e}_{Alight}&0.768 &{.05} &0.022&-0.018& 0 &0\\
\vec{e}_{Aheavy\perp}&&0.891&\bm{0.16} &\bm{-0.14} &0 & 0\\
\vec{e}_{VL} &&&1.13& \bm{-0.354} &0 & 0\\
\vec{e}_{VR}&&&&1.29&0&0\\
\vec{e}_{S}&&&&& 1.00 & 0 \\
\vec{e}_{D}&&&&&& 1.39\\
\hline
\end{array}
$
\caption{ Inner products between the basis vectors $ \vec{e}_i(\LNP\sim m_W) = \bm{G}^*(m_W,m_\m) \hat{e}_i$. On the diagonal are the norms 
(at $m_W$ in SMEFT, for unit-normalised $\hat{e}_i$ at low energy), whereas the off-diagonals are $\cos \eta$, where $\eta$ is the angle 
between the vectors. Overlaps smaller than $10^{-3}$ are given as 0. \label{tab:ip}}
\end{table}

\subsection{A reduced basis at $\LNP \sim m_W$}
\label{sapp:LmW}

In this section we outline a method for choosing a penguin-less basis vector corresponding to a linear combination of $\vec{e }_{VL}$ and $\vec{e }_{VR}$, 
which is approximately orthogonal to the remaining basis vectors. In addition,
the complementarity plots involving $\vec{e }_{VL}$ or $\vec{e }_{VR}$ are similar, so this choice suppresses redundancy.

The dependence of BR($\meee$) on $C_{VL}$ and $C_{VR}$ is very similar, as can be seen from eqn (\ref{BRmeee}). Since these coefficients 
can be distinguished via the angular distributions of the final state electrons in $\meee$~\cite{Okadameee} rather than by comparing BR($\meg$) and BR($\meee$), it is sufficient to plot $\meg$ and $\meee$ in terms of a combination of these coefficients. By choosing this combination to 
suppress penguin contribution, we obtain five approximately orthogonal basis vectors. 

As mentioned above, in the SMEFT basis at $m_W$, $\vec{e }_{VL}$ and $\vec{e }_{VR}$ have components $\sim g^e_L, g^e_R$ along the directions 
of the penguin operators of eqn (\ref{pingouinSMEFT}). So introducing $\vec{e}_{V}\propto g^e_R\vec{e }_{VL} - g^e_L \vec{e }_{VR}$, one sees that
its penguin component vanishes at tree level, and $\vec{e}_{V}$ will be orthogonal to $\vec{e }_{Aheavy\perp}$ up to loop effects. We therefore replace
the $\vec{e}_{VR}\otimes \vec{e}_{VL}$ plane by an axis along $\vec{e}_{V}$
and plot the complementarity of the three rates in the resulting 5-dimensional space.

In this reduced basis, 
the BRs can be written in terms of operator coefficients at $\LNP \sim m_W$ as
\bea
{\rm BR}(\megL) 
 &=& 384\pi^2 \frac{v^4}{\LNP^4} \left[ |\vec{e}_D| \cos\theta_D\right]^{2}
\nonumber\\
 {\rm BR}(\meee) &=& \frac{v^4}{\LNP^4} {\Big [} 
\frac{|\vec{e}_{S}|^2\sin^2\theta_D \cos^2\theta_S}{8}
+ (|\vec{e}^{~'}_{VR}|\sin\theta_D \sin\theta_S \cos\theta_{V}
+ 4e|\vec{e}_D| \cos\theta_D)^2 \nonumber\\&& 
+ 2 (|\vec{e}^{~'}_{VL}|\sin\theta_D \sin\theta_S \cos\theta_{V}
+ 4e|\vec{e}_D| \cos\theta_D)^2+18.76(|\vec{e}_D| \cos\theta_D)^2
{\Big ]} 
\nonumber\\
{\rm BR}(\mu Al \to e Al)
&=& 
\frac{6144 \pi^3v^4}{2.197 g^{cap}_{Al} \LNP^4}
{\Big [}  |\vec{u}_{Al}| |\vec{e}_{Al}|
 \sin\theta_D \sin\theta_S \sin\theta_{V} \sin\phi
 + \frac{D_{Al}}{4}| \vec{e}_{D}| \cos\theta_D {\Big ]} ^{2} 
\nonumber\\
{\rm BR}(\mu Au \to e Au) &=& 
\frac{6144 \pi^3v^4}{2.197 g^{cap}_{Au} \LNP^4}
{\Big [} |\vec{u}_{Au}| ( \cos\theta_{AuAl} |\vec{e}_{Al}|
 \sin\theta_D \sin\theta_S  \sin\theta_{V}\sin\phi
\nonumber\\&&+ \sin\theta_{AuAl} |\vec{e}_{\perp}|
\sin\theta_D \sin\theta_S
\sin\theta_{V}\cos\phi)
 + \frac{D_{Au}}{4}| \vec{e}_{D}| \cos\theta_D {\Big ]} ^{2} 
\label{BRBE}
\eea
where we replace $\{ \theta_{VR},\theta_{VL} \} \to \theta_V$, the un-primed norms of the basis vectors at $m_W$
are on the diagonal of Table~\ref{tab:ip} and $ |\vec{e }^{~'}_{VR}|= -g^e_L|\vec{e }_{VR}| /\sqrt{ (g^e_L)^2+
 (g^e_R)^2}= 1.05$, $|\vec{e }^{~'}_{VL}| = g^e_L|\vec{e }_{VL}|/\sqrt{ (g^e_R)^2+
 (g^e_R)^2}= 0.709 $, 
$\theta_{\rm AuAl}$ is in Table~\ref{tab:misalignment2G}, and norms obtained from the $\muc$ overlap integrals and 
nucleon-quark matching are
$|\vec{u}_{Al}| = 0.39706$
$|\vec{u}_{Ti}| = 0.991275$, and 
$|\vec{u}_{Au}| =  1.92876$.

\subsection{An enlarged basis at $\LNP \sim m_W$}
\label{sapp:LmWps}

This section outlines the approach of adding $Z\to e^\pm \mu ^\mp$ to the observables, and removing the contribution of the 
flavour-changing $Z$ penguin from the four-fermion operators.
This ensures that the basis vectors are orthogonal at $m_W$ to within a degree or two, and highlights the importance of
$\Zme$ for distinguishing among coefficients and models.

The operators ${\cal O}^{e \mu}_{HL1}$ and ${\cal O}^{e \mu}_{HL3}$ of eqn (\ref{pingouinSMEFT}) mediate flavour-changing $Z$ decays, upon 
which ATLAS ~\cite{ATLASZmue} sets the constraint $BR(\Zme)<7.5\times 10^{-7}$ (based on 20 fb$^{-1}$ of luminosity), implying:
\beq
\frac{v^2}{ \LNP^2} | C^{e \mu}_{HL1}+ C^{e \mu}_{HL3}| \lsim 1.6\times 10^{-3}
\label{Zemuexpt}
\eeq
The current sensitivity of $\muc$ and $\meee$ to these coefficients is three orders of magnitude better, and should improve by another
two orders of magnitude with upcoming experiments. Nonetheless, improving the experimental reach in $\Zme$ is interesting, because 
experiments at muon mass scale cannot distinguish these penguin operators from the four-fermion ones\footnote{The situations
of the $Z$-penguin and the dipole are rather different: the dipole is far better constrained than the penguin, because its easier 
to produce muons than $Z$ bosons. However, the dipole is also constrained by $BR(\meee)$, and could be distinguished from four-fermion 
operators using angular distributions in $\meee$, whereas $\Zme$ appears crucial for constraining and identifying the penguins.}

We include an additional basis vector in the coefficient subspace above $m_W$ :
 \bea
 \hat{e}_{ping} = \frac{1}{\sqrt{2}}( \hat{e}_{HL1} + \hat{e}_{HL3)}
\label{eping}
\eea
(for the case of an outgoing $e_L$; for outgoing $e_R$ it would be $\hat{e}_{HE}$), and rewrite the operator coefficients in spherical 
coordinates as
\bea
\vec{C} &=& \frac{1}{\LNP^2}{\Big [}\cos\theta_{ping} \hat{e}_{ping}
+\sin\theta_{ping} \cos\theta_D \hat{e}_{D}
+ \sin\theta_{ping}\sin\theta_D \cos\theta_S \hat{e}_{S,R}
+ \sin\theta_{ping}\sin\theta_D \sin\theta_S \cos\theta_{VR} \hat{e}_{V,R}
\nonumber\\
&&
+ \sin\theta_{ping}\sin\theta_D \sin\theta_S \sin\theta_{VR} \cos\theta_{VL} \hat{e}_{V,L}
 + \sin\theta_{ping}\sin\theta_D \sin\theta_S \sin\theta_{VR} \sin\theta_{VL}
(\sin\phi \hat{e}_{Al} + \cos\phi \hat{e}_\perp) {\Big ]}
~~~~,~~~~ 
\label{angparamping}
\eea
where $\theta_J :0.. \pi$ and $\phi :0.. 2\pi$, and recall that a model would predict the various angles and the scale.

The expressions for low energy BRs in this enlarged basis become more complicated, because the low energy four-fermion coefficients
are expressed as the component from $Z$ exchange, plus the component from four-fermion operator at the weak scale. 
The normalisation of some basis vectors changes as well, becoming:
\bea
 |e_{VL}| = 0.944~,~ |e_{VR}| = 1.041~,~
 |e_{Al}| = 0.768~,~ |e_{Aheavy\perp}| = 0.861 
\eea

The formulae for the reach, obtained from the Branching Ratios are:
\bea
\frac{\Lambda}{v} &=& \left[ \frac{384\pi^2 }{ {\rm BR}(\megL) }\right]^{1/4}
\sqrt{(|\vec{e}_D| \sin\theta_{ping}\cos\theta_D + 1.21\times10^{-3} \cos\theta_{ping})^2}
\label{BRmegping}
\eea
where the one-loop matching of penguin operators to the dipole was included. For $\meee$:
\bea
\frac{\Lambda}{v} &=& \left[\frac{1 }{ {\rm BR}(\meee)}\right]^{1/4}
\sqrt{\frac{|\vec{e}_{S}|\sin\theta_{ping}\sin\theta_D \cos\theta_S}{2\sqrt{2}}}
\nonumber\\
\frac{\Lambda}{v} &=& \left[\frac{1 }{ {\rm BR}(\meee)}\right]^{1/4}
{\Big [} ( |\vec{e}_{VR}|\sin\theta_{ping}\sin\theta_D \sin\theta_S\cos\theta_{VR} + 0.685 \cos\theta_{ping}
+ 4e|\vec{e}_D| \sin\theta_{ping}\cos\theta_D)^2 \nonumber\\ &&
+9.38(|\vec{e}_D| \sin\theta_{ping}\cos\theta_D)^2 {\Big ]} ^{1/4}
\nonumber\\
\frac{\Lambda}{v} &=& \left[\frac{1 }{ {\rm BR}(\meee)}\right]^{1/4}
{\Big [} 2 (|\vec{e}_{VL}|\sin\theta_{ping}\sin\theta_D \sin\theta_S\sin\theta_{VR}
\cos\theta_{VL}
 -0.722 \cos\theta_{ping}
+ 4e|\vec{e}_D| \sin\theta_{ping}\cos\theta_D)^2\nonumber\\ &&+9.38(|\vec{e}_D| \sin\theta_{ping}\cos\theta_D)^2
{\Big ]} ^{1/4} 
\label{BRmucpig}
\eea
where $0.685$ and -$0.722$ correspond respectively to $\sqrt{2} g^e_{R,L}\times$QED loop corrections. Finally, subtracting the 
penguins from the $u$ and $d$ vector operators in $\muc$ gives
\bea
\frac{\Lambda}{v} &=& \left[\frac{ 1.2292\times10^5}{ {\rm BR}(\mu Al \to e Al)}\right]^{1/4}
{\Big [}  |\vec{u}_{Al}| ( |\vec{e}_{Al}|
 \sin\theta_{ping}\sin\theta_D \sin\theta_S\sin\theta_{VR} \sin\theta_{VL}\sin\phi
 -0.0263\cos\theta_{ping}) \nonumber\\&&+
\frac{D_{Al}}{4}| \vec{e}_{D}| \sin\theta_{ping} \cos\theta_D {\Big ]} ^{1/2} 
\label{coderoot}\\
\frac{\Lambda}{v} &=& \left[\frac{6.63435\times 10^3}{ {\rm BR}(\mu Au \to e Au)}\right]^{1/4}
{\Big [} |\vec{u}_{Au}| ( \cos\theta_{\rm AuAl} |\vec{e}_{Al}|
\sin\theta_{ping}\sin\theta_D \sin\theta_S\sin\theta_{VR} \sin\theta_{VL}\sin\phi
\nonumber\\&&+ \sin\theta_{\rm AuAl} |\vec{e}_{\perp}|
\sin\theta_{ping}\sin\theta_D \sin\theta_S\sin\theta_{VR} \sin\theta_{VL}\cos\phi -0.0474\cos\theta_{ping})
 + \frac{D_{Au}}{4}| \vec{e}_{D}| \sin\theta_{ping} \cos\theta_D {\Big ]} ^{1/2} 
\nonumber
\eea

\subsection{The eigenbasis of the covariance matrix}
\label{app:V}

An alternative basis for the subspace of constrained coefficients, also orthogonal and perhaps more familiar, would be the eigenvectors of
the covariance matrix. The inverse covariance matrix $\bm{V}$ for all the processes can be written as
\bea
\vec{C}^\dagger \bm{V} \vec{C}&=& \vec{C}^\dagger \frac{\bm {R}_{\meg} }{B_{\meg}^{expt}} \vec{C}+
\vec{C}^\dagger \frac{\bm {R}_{\meee} }{B_{\meee}^{expt}} \vec{C}+
\vec{C}^\dagger \frac{\bm {R}_{\mucAl} }{B_{\mucAl}^{expt}} \vec{C}+
\vec{C}^\dagger \frac{\bm {R}_{\mucAu} }{B_{\mucAu}^{expt}} \vec{C} ~~,
\label{V1}
\eea
where the coefficients are evaluated at the experimental scale. The inverse eigenvalues give the allowed range of coefficients in 
the eigenbasis, and the eigenvectors are orthogonal combinations of operators, which correspond to the axes of the allowed ellipse
around the origin in coefficient space. This is a convenient basis for plotting, because the limiting values of each parameter are 
obtained on the axes, so there is no need to do perform scans. However, we prefer the operator basis of eqn (\ref{Lag1}),
because it is simple and directly related to the experimental processes.

A covariance matrix for the coefficients at $\LNP$ can be obtained by substituting eqn (\ref{pt5}) into eqn (\ref{V1}). 
This matrix is large, ($\sim 90 \times 90$), so despite that most of the eigenvalues should vanish, finding the eigenvectors 
of the 12 non-zero eigenvalues would be a numerical exercise which could disconnect the final basis and constraints from the 
input processes. It has the advantage of giving an {\it orthonormal} basis, whose eigenvectors correspond to the axes of the 
allowed ellipse in coefficient space.


\begin{thebibliography}{222222}

\bibitem{KO}
Y.~Kuno and Y.~Okada,
``Muon decay and physics beyond the standard model,''
Rev. Mod. Phys. \textbf{73} (2001), 151-202
doi:10.1103/RevModPhys.73.151
[arXiv:hep-ph/9909265 [hep-ph]].


\bibitem{Calibbi}
L.~Calibbi and G.~Signorelli,
``Charged Lepton Flavour Violation: An Experimental and Theoretical Introduction,''
Riv. Nuovo Cim. \textbf{41} (2018) no.2, 71-174
doi:10.1393/ncr/i2018-10144-0
[arXiv:1709.00294 [hep-ph]].

 
\bibitem{mnu}
M.~C.~Gonzalez-Garcia and Y.~Nir,
``Neutrino Masses and Mixing: Evidence and Implications,''
Rev. Mod. Phys. \textbf{75} (2003), 345-402
doi:10.1103/RevModPhys.75.345
[arXiv:hep-ph/0202058 [hep-ph]].




\bibitem{Y+} 
M.~Fukugita and T.~Yanagida,
``Baryogenesis Without Grand Unification,''
Phys. Lett. B \textbf{174} (1986), 45-47
doi:10.1016/0370-2693(86)91126-3



\bibitem{PRep}
S.~Davidson, E.~Nardi and Y.~Nir,
``Leptogenesis,''
Phys. Rept. \textbf{466} (2008), 105-177
doi:10.1016/j.physrep.2008.06.002
[arXiv:0802.2962 [hep-ph]].




\bibitem{TheMEG:2016wtm}
 A.~M.~Baldini {\it et al.} [MEG Collaboration],
 ``Search for the lepton flavour violating decay $\mu ^+ \rightarrow \mathrm {e}^+ \gamma $ with the full dataset of the MEG experiment,''
 Eur.\ Phys.\ J.\ C {\bf 76} (2016) no.8, 434
 doi:10.1140/epjc/s10052-016-4271-x
 [arXiv:1605.05081 [hep-ex]].


\bibitem{MEGII}
 A.~M.~Baldini {\it et al.} [MEG II Collaboration],
 ``The design of the MEG II experiment,''
 Eur.\ Phys.\ J.\ C {\bf 78} (2018) no.5, 380
 doi:10.1140/epjc/s10052-018-5845-6
 [arXiv:1801.04688 [physics.ins-det]].
%
%
\bibitem{Bellgardt:1987du}
 U.~Bellgardt {\it et al.} [SINDRUM Collaboration],
 ``Search for the Decay mu+ -> e+ e+ e-,''
 Nucl.\ Phys.\ B {\bf 299} (1988) 1.
 doi:10.1016/0550-3213(88)90462-2
 



\bibitem{Mu3e}
 A.~Blondel {\it et al.},
 ``Research Proposal for an Experiment to Search for the Decay $\mu \to eee$,''
 arXiv:1301.6113 [physics.ins-det].
 

 
\bibitem{Bertl:2006up}
 W.~H.~Bertl {\it et al.} [SINDRUM II Collaboration],
 ``A Search for muon to electron conversion in muonic gold,''
 Eur.\ Phys.\ J.\ C {\bf 47} (2006) 337.
 doi:10.1140/epjc/s2006-02582-x
C.~Dohmen {\it et al.} [SINDRUM II Collaboration],
 ``Test of lepton flavor conservation in mu -> e conversion on titanium,''
 Phys.\ Lett.\ B {\bf 317} (1993) 631.


\bibitem{COMET}
 Y.~G.~Cui {\it et al.} [COMET Collaboration],
 ``Conceptual design report for experimental search for lepton flavor violating mu- - e- conversion at sensitivity of 10**(-16) with a slow-extracted bunched proton beam (COMET),''
 KEK-2009-10.
M.~L.~Wong [COMET Collaboration],
 ``Overview of the COMET Phase-I experiment,''
 PoS FPCP {\bf 2015} (2015) 059.

\bibitem{mu2e}
 R.~M.~Carey {\it et al.} [Mu2e Collaboration],
 ``Proposal to search for $\mu^- N \to e^- N$ with a single event sensitivity below $10^{-16}$,''
 FERMILAB-PROPOSAL-0973.

\bibitem{PP}
Y.~ Kuno {\it et al.} (PRISM collaboration), ''An Experimental Search for a 
$\mu N\to e N$ Conversion at Sensitivity of the Order of
$10^{-18}$ with a Highly Intense Muon Source: PRISM'', unpublished, J-PARC LOI, 2006.
 
\bibitem{tau1}
Bernard Aubert et~al.
\newblock {Searches for Lepton Flavor Violation in the Decays tau+-
 ---\ensuremath{>} e+- gamma and tau+- ---\ensuremath{>} mu+- gamma}.
\newblock {\em Phys. Rev. Lett.}, 104:021802, 2010.

\bibitem{belle2t3l}
W.~Altmannshofer et~al.
\newblock {The Belle II Physics Book}.
\newblock {\em PTEP}, 2019(12):123C01, 2019.
\newblock [Erratum: PTEP 2020, 029201 (2020)].

\bibitem{tau2}
K.~Hayasaka et~al.
\newblock {Search for Lepton Flavor Violating Tau Decays into Three Leptons
 with 719 Million Produced Tau+Tau- Pairs}.
\newblock {\em Phys. Lett. B}, 687:139--143, 2010.

\bibitem{Belle:2007cio}
Y.~Miyazaki et~al.
\newblock {Search for lepton flavor violating tau- decays into l- eta, l-
 eta-prime and l- pi0}.
\newblock {\em Phys. Lett. B}, 648:341--350, 2007.


\bibitem{everyone}

J.~Hisano, T.~Moroi, K.~Tobe and M.~Yamaguchi,
``Lepton flavor violation via right-handed neutrino Yukawa couplings in supersymmetric standard model,''
Phys. Rev. D \textbf{53} (1996), 2442-2459
doi:10.1103/PhysRevD.53.2442
[arXiv:hep-ph/9510309 [hep-ph]].
 
W.~Altmannshofer, A.~J.~Buras, S.~Gori, P.~Paradisi and D.~M.~Straub,
``Anatomy and Phenomenology of FCNC and CPV Effects in SUSY Theories,''
Nucl. Phys. B \textbf{830} (2010), 17-94
doi:10.1016/j.nuclphysb.2009.12.019
[arXiv:0909.1333 [hep-ph]].

M.~Blanke, A.~J.~Buras, B.~Duling, A.~Poschenrieder and C.~Tarantino,
``Charged Lepton Flavour Violation and (g-2)(mu) in the Littlest Higgs Model with T-Parity: A Clear Distinction from Supersymmetry,''
JHEP \textbf{05} (2007), 013
doi:10.1088/1126-6708/2007/05/013
[arXiv:hep-ph/0702136 [hep-ph]].

Y.~Omura, E.~Senaha and K.~Tobe,
``Lepton-flavor-violating Higgs decay $h \to \mu\tau$ and muon anomalous magnetic moment in a general two Higgs doublet model,''
JHEP \textbf{05} (2015), 028
doi:10.1007/JHEP05(2015)028
[arXiv:1502.07824 [hep-ph]].

E.~Arganda, M.~J.~Herrero, X.~Marcano and C.~Weiland,
``Imprints of massive inverse seesaw model neutrinos in lepton flavor violating Higgs boson decays,''
Phys. Rev. D \textbf{91} (2015) no.1, 015001
doi:10.1103/PhysRevD.91.015001
[arXiv:1405.4300 [hep-ph]].


F.~Deppisch and J.~W.~F.~Valle,
``Enhanced lepton flavor violation in the supersymmetric inverse seesaw model,''
Phys. Rev. D \textbf{72} (2005), 036001
doi:10.1103/PhysRevD.72.036001
[arXiv:hep-ph/0406040 [hep-ph]].



S.~Antusch, E.~Arganda, M.~J.~Herrero and A.~M.~Teixeira,
``Impact of theta(13) on lepton flavour violating processes within SUSY seesaw,''
JHEP \textbf{11} (2006), 090
doi:10.1088/1126-6708/2006/11/090
[arXiv:hep-ph/0607263 [hep-ph]].


T.~M.~Aliev, A.~S.~Cornell and N.~Gaur,
``Lepton flavour violation in unparticle physics,''
Phys. Lett. B \textbf{657} (2007), 77-80
doi:10.1016/j.physletb.2007.09.055
[arXiv:0705.1326 [hep-ph]].



M.~L.~L\'opez-Ib\'a\~nez, A.~Melis, M.~J.~P\'erez, M.~H.~Rahat and O.~Vives,
``Constraining low-scale flavor models with (g-2)\ensuremath{\mu} and lepton flavor violation,''
Phys. Rev. D \textbf{105} (2022) no.3, 035021
doi:10.1103/PhysRevD.105.035021
[arXiv:2112.11455 [hep-ph]].


P.~Escribano, M.~Hirsch, J.~Nava and A.~Vicente,
``Observable flavor violation from spontaneous lepton number breaking,''
JHEP \textbf{01} (2022), 098
doi:10.1007/JHEP01(2022)098
[arXiv:2108.01101 [hep-ph]].

Y.~Cai, J.~Herrero-Garc\'\i{}a, M.~A.~Schmidt, A.~Vicente and R.~R.~Volkas,
``From the trees to the forest: a review of radiative neutrino mass models,''
Front. in Phys. \textbf{5} (2017), 63
doi:10.3389/fphy.2017.00063
[arXiv:1706.08524 [hep-ph]].

\bibitem{deGouvea}
A.~de Gouvea and P.~Vogel,
``Lepton Flavor and Number Conservation, and Physics Beyond the Standard Model,''
Prog. Part. Nucl. Phys. \textbf{71} (2013), 75-92
doi:10.1016/j.ppnp.2013.03.006
[arXiv:1303.4097 [hep-ph]].

\bibitem{PSI}
A.~Crivellin, S.~Davidson, G.~M.~Pruna and A.~Signer,
``Renormalisation-group improved analysis of $\mu\to e$ processes in a systematic effective-field-theory approach,''
JHEP \textbf{05} (2017), 117
[arXiv:1702.03020 [hep-ph]].

\bibitem{C+C}
S.~Davidson,
``Completeness and complementarity for $\mu \to e\gamma \mu \to e \bar e e$ and $\mu A \to eA$,''
JHEP \textbf{02} (2021), 172
doi:10.1007/JHEP02(2021)172
[arXiv:2010.00317 [hep-ph]].



  

\bibitem{Georgi} 
 H.~Georgi,
 ``Effective field theory,''
 Ann.\ Rev.\ Nucl.\ Part.\ Sci.\ {\bf 43 } (1993) 209-252.\\
 H.~Georgi,
 ``On-shell effective field theory,''
 Nucl.\ Phys.\ {\bf B361 } (1991) 339-350.
 
\bibitem{burashouches}
 A.~J.~Buras,
 ``Weak Hamiltonian, CP violation and rare decays,''
 hep-ph/9806471.


\bibitem{LesHouches}
Les Houches Lect. Notes \textbf{108} (2020).
 A.~V.~Manohar,
``Introduction to Effective Field Theories,''
[arXiv:1804.05863 [hep-ph]].

A.~Pich,
``Effective Field Theory with Nambu-Goldstone Modes,''
[arXiv:1804.05664 [hep-ph]].
 
L.~Silvestrini,
``Effective Theories for Quark Flavour Physics,''
[arXiv:1905.00798 [hep-ph]].
 
 M.~Balsiger, M.~Bounakis, M.~Drissi, J.~Gargalionis, E.~Gustafson, G.~Jackson, M.~Leak, C.~Lepenik, S.~Melville and D.~Moreno, \textit{et al.}
 ``Solutions to Problems at Les Houches Summer School on EFT,''
[arXiv:2005.08573 [hep-ph]].

\bibitem{DKY}
 S.~Davidson, Y.~Kuno and M.~Yamanaka,
 ``Selecting $\mu \to e$ conversion targets to distinguish lepton flavour-changing operators,''
 Phys.\ Lett.\ B {\bf 790} (2019) 380
 doi:10.1016/j.physletb.2019.01.042
 [arXiv:1810.01884 [hep-ph]].

\bibitem{JMT} 
 R.~Alonso, E.~E.~Jenkins, A.~V.~Manohar and M.~Trott,
 ``Renormalization Group Evolution of the Standard Model Dimension Six Operators III: Gauge Coupling Dependence and Phenomenology,''
 JHEP {\bf 1404} (2014) 159
 [arXiv:1312.2014 [hep-ph]].
 E.~E.~Jenkins, A.~V.~Manohar and M.~Trott,
 ``Renormalization Group Evolution of the Standard Model Dimension Six Operators II: Yukawa Dependence,''
 JHEP {\bf 1401} (2014) 035
 doi:10.1007/JHEP01(2014)035
 [arXiv:1310.4838 [hep-ph]].
 


\bibitem{CDK}
 V.~Cirigliano, S.~Davidson and Y.~Kuno,
 ``Spin-dependent $\mu \to e$ conversion,''
 Phys.\ Lett.\ B {\bf 771} (2017) 242
 doi:10.1016/j.physletb.2017.05.053
 [arXiv:1703.02057 [hep-ph]].


\bibitem{DKS}
 S.~Davidson, Y.~Kuno and A.~Saporta,
 ``Spin-dependent ${\mu \rightarrow e}$ conversion on light nuclei,''
 Eur.\ Phys.\ J.\ C {\bf 78} (2018) no.2, 109
 doi:10.1140/epjc/s10052-018-5584-8
 [arXiv:1710.06787 [hep-ph]].

 
\bibitem{Roma} 
 M.~Ciuchini, E.~Franco, L.~Reina and L.~Silvestrini,
 ``Leading order QCD corrections to b ---> s gamma and b ---> s g decays in three regularization schemes,''
 Nucl.\ Phys.\ B {\bf 421} (1994) 41
 [hep-ph/9311357].

\bibitem{Murphy} 
Christopher~W. Murphy.
\newblock {Dimension-8 operators in the Standard Model Eective Field Theory}.
\newblock {\em JHEP}, 10:174, 2020.

\bibitem{autredim8}
Hao-Lin Li, Zhe Ren, Jing Shu, Ming-Lei Xiao, Jiang-Hao Yu, and Yu-Hui Zheng.
\newblock {Complete set of dimension-eight operators in the standard model
 effective field theory}.
\newblock {\em Phys. Rev. D}, 104(1):015026, 2021.


\bibitem{powercount} 
M.~Ardu and S.~Davidson,
``What is Leading Order for LFV in SMEFT?,''
JHEP \textbf{08} (2021), 002
doi:10.1007/JHEP08(2021)002
[arXiv:2103.07212 [hep-ph]].



\bibitem{sphcoord}
 L. E. Blumenson,
``A Derivation of n-Dimensional Spherical Coordinates'',
 The American Mathematical Monthly, Vol. 67, No. 1 (Jan., 1960), pp. 63-66
 http://www.jstor.org/stable/2308932

 \bibitem{Okadameee}
 Y.~Okada, K.~i.~Okumura and Y.~Shimizu,
 ``Mu --> e gamma and mu --> 3 e processes with polarized muons and supersymmetric grand unified theories,''
 Phys.\ Rev.\ D {\bf 61} (2000) 094001
 doi:10.1103/PhysRevD.61.094001
 [hep-ph/9906446].

Y.~Okada, K.~i.~Okumura and Y.~Shimizu,
 ``CP violation in the mu ---> 3 e process and supersymmetric grand unified theory,''
 Phys.\ Rev.\ D {\bf 58} (1998) 051901
 doi:10.1103/PhysRevD.58.051901
 [hep-ph/9708446].



\bibitem{Hisano:1996qq}
J.~Hisano, T.~Moroi, K.~Tobe and M.~Yamaguchi,
``Exact event rates of lepton flavor violating processes in supersymmetric SU(5) model,''
Phys. Lett. B \textbf{391} (1997), 341-350
[erratum: Phys. Lett. B \textbf{397} (1997), 357]
doi:10.1016/S0370-2693(96)01473-6
[arXiv:hep-ph/9605296 [hep-ph]].




\bibitem{DBC}
S.~Davidson, D.~C.~Bailey and B.~A.~Campbell,
``Model independent constraints on leptoquarks from rare processes,''
Z. Phys. C \textbf{61} (1994), 613-644
doi:10.1007/BF01552629
[arXiv:hep-ph/9309310 [hep-ph]].


\bibitem{Slovene}
I.~Dor\v{s}ner, S.~Fajfer, A.~Greljo, J.~F.~Kamenik and N.~Ko\v{s}nik,
``Physics of leptoquarks in precision experiments and at particle colliders,''
Phys. Rept. \textbf{641} (2016), 1-68
doi:10.1016/j.physrep.2016.06.001
[arXiv:1603.04993 [hep-ph]].


\bibitem{Hetal}
 V.~Cirigliano, M.~L.~Graesser and G.~Ovanesyan,
``WIMP-nucleus scattering in chiral effective theory,''
JHEP \textbf{10} (2012), 025
doi:10.1007/JHEP10(2012)025
[arXiv:1205.2695 [hep-ph]].

M.~Hoferichter, P.~Klos and A.~Schwenk,
``Chiral power counting of one- and two-body currents in direct detection of dark matter,''
Phys. Lett. B \textbf{746} (2015), 410-416
doi:10.1016/j.physletb.2015.05.041
[arXiv:1503.04811 [hep-ph]].



\bibitem{Crivellin:2014cta}
A.~Crivellin, M.~Hoferichter and M.~Procura,
``Improved predictions for $\mu\to e$ conversion in nuclei and Higgs-induced lepton flavor violation,''
Phys. Rev. D \textbf{89} (2014), 093024
doi:10.1103/PhysRevD.89.093024
[arXiv:1404.7134 [hep-ph]].


\bibitem{VCetal}
V.~Cirigliano, K.~Fuyuto, M.~J.~Ramsey-Musolf and E.~Rule,
``Next-to-leading order scalar contributions to $\mu\rightarrow e$ conversion,''
[arXiv:2203.09547 [hep-ph]].


\bibitem{Dekens}
W.~Dekens, E.~E.~Jenkins, A.~V.~Manohar and P.~Stoffer,
``Non-perturbative effects in $\mu \to e \gamma$,''
JHEP \textbf{01} (2019), 088
doi:10.1007/JHEP01(2019)088
[arXiv:1810.05675 [hep-ph]].

\bibitem{KKO}
 R.~Kitano, M.~Koike and Y.~Okada,
 ``Detailed calculation of lepton flavor violating muon electron conversion rate for various nuclei,''
 Phys.\ Rev.\ D {\bf 66} (2002) 096002
  Erratum: [Phys.\ Rev.\ D {\bf 76} (2007) 059902]
 doi:10.1103/PhysRevD.76.059902, 10.1103/PhysRevD.66.096002
 [hep-ph/0203110].

\bibitem{LLF}
S.~Borsanyi, Z.~Fodor, C.~Hoelbling, L.~Lellouch, K.~K.~Szabo, C.~Torrero and L.~Varnhorst,
``Ab-initio calculation of the proton and the neutron's scalar couplings for new physics searches,''
[arXiv:2007.03319 [hep-lat]].


\bibitem{Hoferichter:2015dsa}
 M.~Hoferichter, J.~Ruiz de Elvira, B.~Kubis and U.~G.~Meissner,
 ``High-Precision Determination of the Pion-Nucleon Term from Roy-Steiner Equations,''
Phys.\ Rev.\ Lett.\ {\bf 115} (2015) 092301
doi:10.1103/PhysRevLett.115.092301
[arXiv:1506.04142 [hep-ph]].

%
\bibitem{Crivellin:2013ipa}
A.~Crivellin, M.~Hoferichter and M.~Procura,
``Accurate evaluation of hadronic uncertainties in spin-independent WIMP-nucleon scattering: Disentangling two- and three-flavor effects,''
Phys. Rev. D \textbf{89} (2014), 054021
doi:10.1103/PhysRevD.89.054021
[arXiv:1312.4951 [hep-ph]].

%
\bibitem{RuizdeElvira:2017stg}
J.~Ruiz de Elvira, M.~Hoferichter, B.~Kubis and U.~G.~Mei\ss{}ner,
``Extracting the $\sigma$-term from low-energy pion-nucleon scattering,''
J. Phys. G \textbf{45} (2018) no.2, 024001
doi:10.1088/1361-6471/aa9422
[arXiv:1706.01465 [hep-ph]].


\bibitem{Lellouch}
 S.~Durr {\it et al.},
 ``Lattice computation of the nucleon scalar quark contents at the physical point,''
 Phys.\ Rev.\ Lett.\ {\bf 116} (2016) no.17, 172001
 doi:10.1103/PhysRevLett.116.172001
 [arXiv:1510.08013 [hep-lat]].


\bibitem{Junnarkar:2013ac}
 P.~Junnarkar and A.~Walker-Loud,
 ``Scalar strange content of the nucleon from lattice QCD,''
Phys.\ Rev.\ D {\bf 87} (2013) 114510
doi:10.1103/PhysRevD.87.114510
[arXiv:1301.1114 [hep-lat]].



\bibitem{Alarcon:2011zs}
 J.~M.~Alarcon, J.~Martin Camalich and J.~A.~Oller,
 ``The chiral representation of the $\pi N$ scattering amplitude and the pion-nucleon sigma term,''
 Phys.\ Rev.\ D {\bf 85} (2012) 051503
 doi:10.1103/PhysRevD.85.051503
 [arXiv:1110.3797 [hep-ph]].


%
\bibitem{Gupta:2021ahb}
R.~Gupta, S.~Park, M.~Hoferichter, E.~Mereghetti, B.~Yoon and T.~Bhattacharya,
``Pion\textendash{}Nucleon Sigma Term from Lattice QCD,''
Phys. Rev. Lett. \textbf{127} (2021) no.24, 24
doi:10.1103/PhysRevLett.127.242002
[arXiv:2105.12095 [hep-lat]].


 
\bibitem{SVZ}
 M.~A.~Shifman, A.~I.~Vainshtein and V.~I.~Zakharov,
 ``Remarks on Higgs Boson Interactions with Nucleons,''
 Phys.\ Lett.\ B {\bf 78} (1978) 443.




\bibitem{Agashe:2014kda}
 K.~A.~Olive {\it et al.} [Particle Data Group],
 ``Review of Particle Physics,''
 Chin.\ Phys.\ C {\bf 38} (2014) 090001.
 doi:10.1088/1674-1137/38/9/090001



\bibitem{PDB}
P.~A.~Zyla \textit{et al.} [Particle Data Group],
``Review of Particle Physics,''
PTEP \textbf{2020} (2020) no.8, 083C01






\bibitem{DKUY}
S.~Davidson, Y.~Kuno, Y.~Uesaka and M.~Yamanaka,
``Probing $\mu e \gamma \gamma$ contact interactions with $\mu \to e$ conversion,''
Phys. Rev. D \textbf{102} (2020) no.11, 115043
[arXiv:2007.09612 [hep-ph]].


 
\bibitem{Suzuki:1987jf}
 T.~Suzuki, D.~F.~Measday and J.~P.~Roalsvig,
 ``Total Nuclear Capture Rates for Negative Muons,''
 Phys.\ Rev.\ C {\bf 35} (1987) 2212.
 doi:10.1103/PhysRevC.35.2212



\bibitem{CKOT}
 V.~Cirigliano, R.~Kitano, Y.~Okada and P.~Tuzon,
 ``On the model discriminating power of mu ---> e conversion in nuclei,''
 Phys.\ Rev.\ D {\bf 80} (2009) 013002
 doi:10.1103/PhysRevD.80.013002
 [arXiv:0904.0957 [hep-ph]].


 
 
\bibitem{polonais}
 W.~Buchmuller and D.~Wyler,
 ``Effective Lagrangian Analysis of New Interactions and Flavor Conservation,''
 Nucl.\ Phys.\ B {\bf 268} (1986) 621.
 doi:10.1016/0550-3213(86)90262-2

B.~Grzadkowski, M.~Iskrzynski, M.~Misiak and J.~Rosiek,
 ``Dimension-Six Terms in the Standard Model Lagrangian,''
 JHEP {\bf 1010} (2010) 085
 [arXiv:1008.4884 [hep-ph]].


 
\bibitem{ATLASZmue}
G.~Aad \textit{et al.} [ATLAS],
``Search for the lepton flavor violating decay
$Z\rightarrow e\bar{\mu}$ in pp collisions at $\sqrt{s}$ TeV with the ATLAS detector,''
Phys. Rev. D \textbf{90} (2014) no.7, 072010
doi:10.1103/PhysRevD.90.072010
[arXiv:1408.5774 [hep-ex]].





\end{thebibliography}
\end{document}